\documentclass[aps,prb, superscriptaddress,
 amsmath,amssymb,
 reprint,%
]{revtex4-2}

\usepackage{graphicx}
\usepackage{subfigure}
\usepackage{dcolumn}
\usepackage{multirow}
\usepackage{bm}
\usepackage{array}
\usepackage[colorlinks=true,linkcolor=blue,anchorcolor=red,citecolor=blue,urlcolor=black]{hyperref}
\usepackage{comment}

\usepackage{float}
\usepackage{longtable}
\usepackage{mathptmx}
\usepackage{etoolbox, xcolor}
\usepackage{soul}
\usepackage{amssymb}
\usepackage{pifont}

\makeatletter
\def\ignorecitefornumbering#1{%
     \begingroup
         \@fileswfalse
         #1
    \endgroup
}
\makeatother

\begin{document}

\title{MA$_2$Z$_4$ Family Heterostructures: Promises and Prospects}

\author{Che Chen Tho}
\affiliation{ 
Science, Mathematics and Technology (SMT), Singapore University of Technology and Design (SUTD), Singapore 487372}%

\author{San-Dong Guo}
\affiliation{School of Electronic Engineering, Xi’an University of Posts and Telecommunications, Xi’an 710121, China}

\author{Shi-Jun Liang}
\affiliation{National Laboratory of Solid State Microstructures, School of Physics, Collaborative Innovation Center of Advanced Microstructures, Nanjing University, Nanjing 210093, China}

\author{Wee Liat Ong}
\affiliation{ZJU-UIUC Institute, College of Energy Engineering, Zhejiang University, Jiaxing, Haining, Zhejiang, 314400, China}
\affiliation{State Key Laboratory of Clean Energy Utilization, Zhejiang University, Hangzhou, Zhejiang, 310027, China}

\author{\\Chit Siong Lau}
\affiliation{Institute of Materials Research and Engineering (IMRE), Agency for Science, Technology and Research (A*STAR), Singapore 138634, Republic of Singapore}

\author{Liemao Cao}
\affiliation{College of Physics and Electronic Engineering, Hengyang Normal University, Hengyang 421002, China}

\author{Guangzhao Wang}%
\email{wangyan6930@yznu.edu.cn}
\affiliation{Key Laboratory of Extraordinary Bond Engineering and Advanced Materials Technology of Chongqing, School of Electronic Information Engineering, Yangtze Normal University, Chongqing 408100, People’s Republic of China}%

\author{Yee Sin Ang}
\email{yeesin\_ang@sutd.edu.sg}
\affiliation{ 
Science, Mathematics and Technology (SMT), Singapore University of Technology and Design (SUTD), Singapore 487372}%

\begin{abstract}
Recent experimental synthesis of ambient-stable $\mathrm{MoSi_2N_4}$ monolayer have garnered enormous research interests. The intercalation morphology of $\mathrm{MoSi_2N_4}$ -- composed of a transition metal nitride (Mo-N) inner sub-monolayer sandwiched by two silicon nitride (Si-N) outer sub-monolayers -- have motivated the computational discovery of an expansive family of \emph{synthetic} $\mathrm{MA_2Z_4}$ monolayers with no bulk (3D) material counterpart (where M = transition metals or alkaline earth metals; A = Si, Ge; and N = N, P, As). $\mathrm{MA_2Z_4}$ monolayers exhibit interesting electronic, magnetic, optical, spintronic, valleytronic and topological properties, making them a compelling material platform for next-generation device technologies. Furthermore, heterostructure engineering enormously expands the opportunities of $\mathrm{MA_2Z_4}$. In this review, we summarize the recent rapid progress in the computational design of $\mathrm{MA_2Z_4}$-based heterostructures based on first-principle density functional theory (DFT) simulations - a central \emph{work horse} widely used to understand the physics, chemistry and general design rules for specific targeted functions. We systematically classify the $\mathrm{MA_2Z_4}$-based heterostructures based on their contact types, and review their physical properties, with a focus on their performances in electronics, optoelectronics and energy conversion applications. We review the performance and promises of $\mathrm{MA_2Z_4}$-based heterostructures for device applications that include electrical contacts, transistors, spintronic devices, photodetectors, solar cells, and photocatalytic water splitting. We present several prospects for the computational design of $\mathrm{MA_2Z_4}$-based heterostructures, which hold the potential to guide the next phase of exploration, moving beyond the initial 'gold rush' of $\mathrm{MA_2Z_4}$ research.
This review unveils the vast device application potential of $\mathrm{MA_2Z_4}$-based heterostructures, and paves a roadmap for the future development of $\mathrm{MA_2Z_4}$-based functional heterostructures and devices.         
\end{abstract}

\maketitle

\section{\label{sec:introduction}Introduction}

Two-dimensional (2D) materials offer an exciting platform for next-generation device and renewable energy technology, such as ultrascaled transistors \cite{Quhe2021}, high-efficiency solar cells \cite{Wang2018_photovoltaic}, catalysts \cite{Wang2022_catalyst}, energy storage systems \cite{Zhang2016_storage}, neuromorphic devices \cite{Huh2020_memristors}, valleytronics \cite{Goh2023_book, Vitale2018, Schaibley2016, val1, val2} and quantum technology \cite{Liu2019_information}. The family of 2D materials, such as transition metal dichalcogenides \cite{Choi2017}, phosphorene \cite{Sang2019_phosphorene}, and elemental monolayers \cite{Zhang2016_monolayers}, have been continually expanding since the past decade, which greatly enriches the device applications potential of 2D materials and concretely establishes 2D materials as compelling building blocks for a large array of future solid-state device technology. 

\begin{figure*}
\centering
\includegraphics[width=1\textwidth]{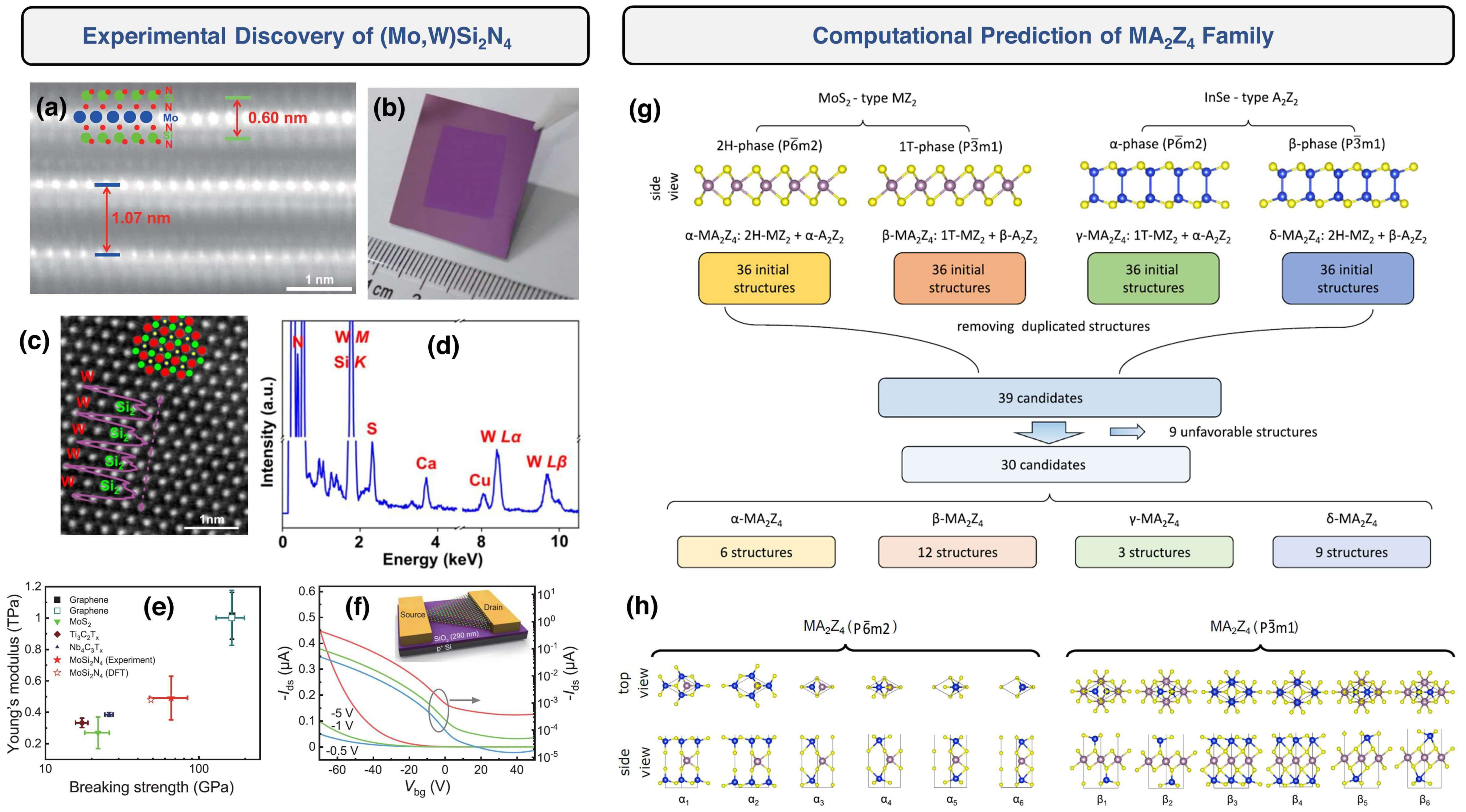}
\caption{\label{Fig1}\textbf{Overview of MoSi$_2$N$_4$ and the MA$_2$Z$_4$ monolayer family}. Experimental synthesis of MoSi$_2$N$_4$ and WSi$_2$N$_4$, showing (a) Cross-sectional scanning transmission electron microscopy image of multilayer MoSi$_2$N$_4$; (b) CVD-grown thin film  of MoSi$_2$N$_4$ on a SiO$_2$/Si substrate. (c) Out-of-plane view of $\mathrm{WSi_2N_4}$. Red, green and yellow dots represent W, Si and N atoms, respectively. Purple curve denotes the intensity profile along the purple dashed line. (d) Energy dispersive x-ray spectroscopy (EDS) profile of $\mathrm{WSi_2N_4}$ showing the W, Si and N atoms from $\mathrm{WSi_2N_4}$, together with Cu, Ca and S atoms which are residues from etching and growth processes. (e) Young’s modulus and breaking strength of monolayer MoSi$_2$N$_4$, monolayer graphene, Mo$S_2$ and MXenes. (f) $\mathrm{MoSi_2N_4}$ transistor and the transfer characteristic at 77 K. (g) Schematic depicting the intercalation of $\mathrm{MoS_2}$ and $\mathrm{InSe}$-liked monolayers to form $\mathrm{MA_2Z_4}$ monolayer. (h) Top and side views of the unit cells of various $\alpha$ and $\beta$ phases. (a)-(f) Reproduced with permission from Hong $et \ al.$, Science \textbf{369}, 670 (2020). Copyright 2020 American Association for the Advancement of Science.
 (g) Reproduced with permission from Wang $et \ al.$, Nat. Commun. \textbf{12}, 2361 (2021). Copyright 2021 Author(s); licensed under a Creative Commons Attribution (CC BY) license.}
\end{figure*}

Recently, 2D semiconducting monolayers MoSi$_2$N$_4$ and WSi$_2$N$_4$ have been synthesized experimentally \cite{Hong2020} via chemical vapor deposition (CVD) and characterized [see Fig. \ref{Fig1}(a)-(d)].
Such monolayers are composed of unstable transition metal nitride cores passivated by two sandwiching Si-N outer-layers.
The MoSi$_2$N$_4$ and WSi$_2$N$_4$ monolayers represent an unusual septuple-atomic-layered monolayers that are synthesized exclusively via `bottom-up' approach since they have no 3D counterparts \cite{Hong2020}.
Importantly, $\mathrm{MoSi_2N_4}$ monolayer exhibits excellent mechanical properties, with Young's modulus (491.4$\pm$139.1 GPa) and tensile strength (65.8$\pm$18.3 GPa) higher than that of Mxenes and $\mathrm{MoS_2}$ [see Fig. \ref{Fig1}(e)]. The optical transmittance of $\mathrm{MoSi_2N_4}$ is comparable to that of graphene.
Furthermore, $\mathrm{MoSi_2N_4}$ can be used as $p$-type channel material for field-effect transistor (FET) $p$-type transistor applications, exhibiting a high on/off ratio of $\sim$4000 at 77K [see Fig. \ref{Fig1}(f) for the transfer characteristic]. Together with its exceptional ambient stability, optical \cite{bafekry2021mosi2n4} and excitonic \cite{kong2022comprehensive} properties, $\mathrm{MoSi_2N_4}$ is a promising 2D material for electronics and optoelectronics device applications.


The morphology of MoSi$_2$N$_4$ and WSi$_2$N$_4$ can be generalized into an expansive family of MA$_2$Z$_4$ monolayers.
A high throughput computational screening exploring the intercalation of an MoS$_2$-liked MZ$_2$ monolayer in 1T or 2H phase with two InSe-liked A$_2$Z$_2$ outer monolayers in $\alpha$ or $\beta$ phase [see Fig. \ref{Fig1}(g)] (M = ealy transition metal, alkali-earth metal, VIB element; A = Al, Ga, Si, Ge; Z = S, Se, Te, N, P, As) predicts 72 thermodynamically and dynamically stable $\mathrm{MA_2Z_4}$ monolayers \cite{Wang2021_MA2Z4}. 
Structurally, the intercalation morphology of $\mathrm{MA_2Z_4}$ enables four phases ($\alpha$, $\beta$, $\delta$, $\gamma$) with each phase consisting of several substructures [see Fig. \ref{Fig1}(h)]. Typically, the $\alpha_1$, $\alpha_2$, $\beta_1$, $\beta_2$ structures are the most energetically stable, and their electronic properties can be classified according to the number of valence electrons (i.e. 32, 33 or 34), which gives rise to rich phenomena, such as ferromagnetism, nontrivial band topologies, valleytronic properties and superconductivity \cite{Wang2021_MA2Z4}.

Following the experimental discovery of MoSi$_2$N$_4$ and WSi$_2$N$_4$ monolayers and the computational high-throughput discovery of the broader $\mathrm{MA_{2}Z_{4}}$ family, intensive theoretical and computational efforts have been devoted to investigate the physical properties and the potential device application of $\mathrm{MA_{2}Z_{4}}$. 
As it is impossible for experimental probes to explore the family of $\mathrm{MA_{2}Z_{4}}$, predictive materials modelling via DFT provides a powerful tool to identify $\mathrm{MA_{2}Z_{4}}$ that are promising for specific device applications.
The proof-of-concept demonstrations of applications, such as transistors \cite{Zhao2021_transistor, Sun2021_transistor, Huang2021_transistor, Ye2022_transistor, Nandan2021_transistor, arxiv_trans}, contacts \cite{Pham2021}, photodetectors \cite{Shu2022, Tho2022}, solar cells \cite{Guo2022_BP,Liu2022,Nguyen2022_C3N4,JinQuan2022,Bafekry2021}, photocatalysts \cite{Xuefeng2022,Zeng2021,He2022,Mortazavi2021}, luminescence devices \cite{Cai2021_MoSe2,Cai2021_WSe2,JinQuan2022}, thermal management systems \cite{Mortazavi2021}, battery \cite{Li2022_battery}, gas sensors \cite{xiao2022_gas, Bafekry2021_gas} and piezoelectric devices \cite{Zhong2021}, have firmly established $\mathrm{MA_{2}Z_{4}}$ as a potential building block of materials for future device technology. 

\begin{figure*}
\centering
\includegraphics[width=1\textwidth]{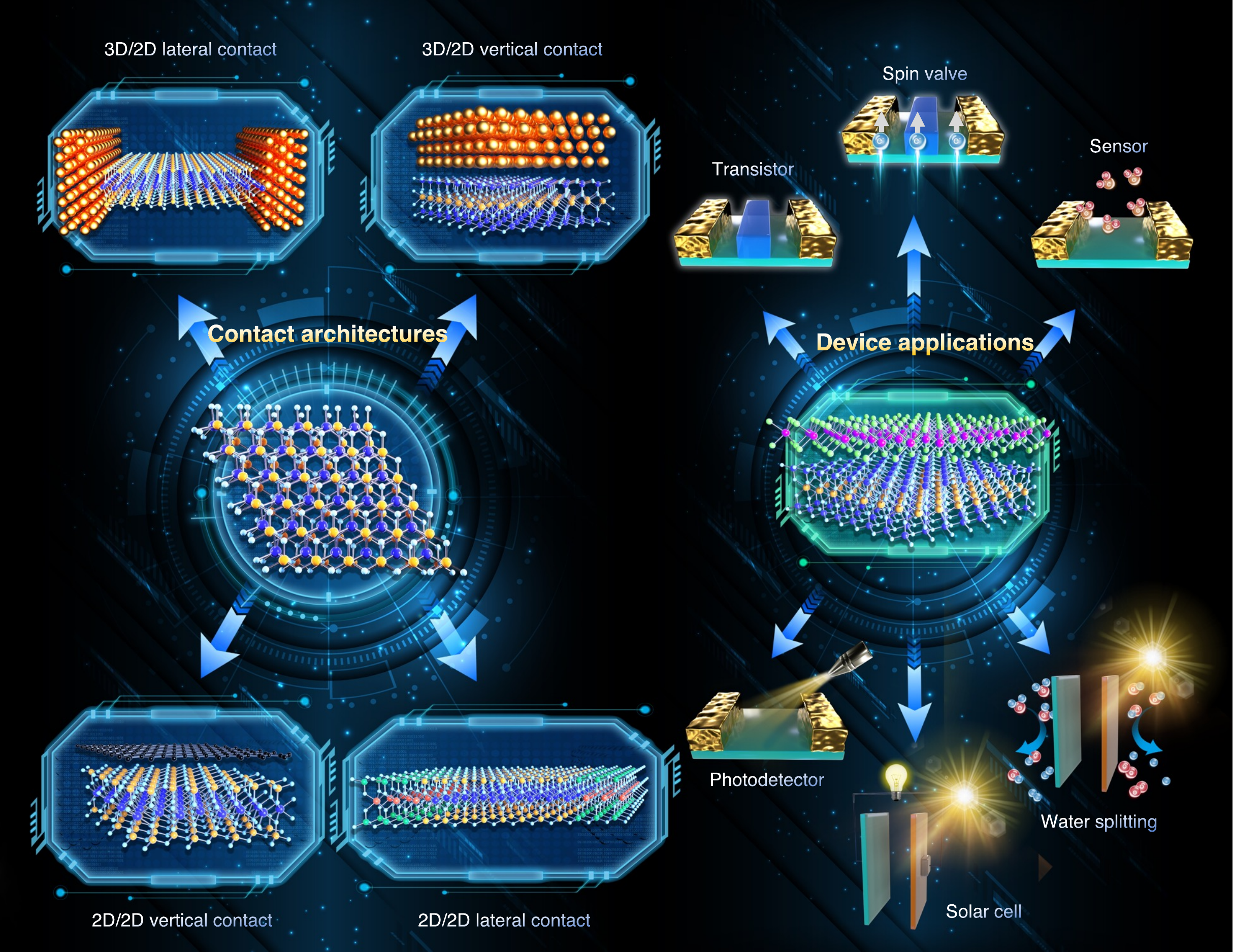}
\caption{\label{Schematic} \textbf{Overview of MA$_2$Z$_4$ heterostructures and applications.} Various contact architectures, including 3D/2D and 2D/2D contacts under both vertical and lateral configurations are schematically shown in the left panel. Representative applications of MA$_2$Z$_4$ heterostructures, such as field-effect transistors, spintronic devices, gas sensors, photodetectors, solar cells and photocatalysts are shown in the right panel.}
\end{figure*}

Beyond monolayers, 2D materials can be stacked vertically to form van der Waals heterostructures (vdWHs) \cite{geim2013van}. Unlike traditional 3D materials, the absence of dangling bonds on 2D material surfaces allows vertical \emph{stacking} via weak interlayer van der Waals (vdW) forces, with minimal strain-induced lattice mismatch and interfacial disorder. Such vertical van der Waals (vdW) stacking architecture further enables the use of rotational degree of freedom between layers to engineer the properties unique to vdWHs \cite{Xin2022_twist}. 
vdWHs thus offer a versatile route for synergizing the physical properties of different 2D materials, which enormously expands the design space and functionalities of 2D material devices \cite{Pham2022}.

Motivated by the enormous progress made in transition metal dichalcogenide (TMDC)-based vdWHs, $\mathrm{MA_{2}Z_{4}}$-based vdWHs and lateral heterostructures have been extensively studied. In contrast to the monolayer forms of $\mathrm{MA_{2}Z_{4}}$ (see Ref. \cite{Yin2022} for a recent review of $\mathrm{MA_{2}Z_{4}}$ monolayers), $\mathrm{MA_{2}Z_{4}}$-based vdWHs possess greater design challenges due to the enormously large possibilities of material combinations. 
A bird's-eye view on the current progress of $\mathrm{MA_{2}Z_{4}}$-based heterostructures, which is needed to tame the enormous design space of $\mathrm{MA_{2}Z_{4}}$-based vdWHs, remains largely incomplete thus far. 
In this work, we provide a comprehensive review on the theoretical and computational studies of $\mathrm{MA_{2}Z_{4}}$-based heterostructures [see Fig. 2 for an overview of $\mathrm{MA_{2}Z_{4}}$-based heterostructures and their applications]. 
The physical and interfacial properties of $\mathrm{MA_{2}Z_{4}}$-based vdWHs as well as their potential applications and device performance characteristics are summarized, with specific focus on electronics, optoelectroncis and renewable energy applications. 
This review aims to provide a summary on the current progress and design challenges of $\mathrm{MA_{2}Z_{4}}$-based vdWHs, and guide the development of novel $\mathrm{MA_{2}Z_{4}}$-based heterostructure device technology.

\textbf{Outline of this review.} 
This review is presented as followed: \textbf{Section \ref{sec:vdWH}} provides an overview of MA$_2$Z$_4$ contact heterostructures; \textbf{Section \ref{MS_Contact}} discuss the general classification and an overview of MA$_2$Z$_4$-based contacts; \textbf{Section \ref{SS Contact}} focuses on the physical properties of MA$_2$Z$_4$-based SS vdWH contacts; \textbf{Section \ref{Beyond_Heterobilayer}} summarizes the properties of MA$_2$Z$_4$-based heterostructures beyond bilayer vdWHs, such as lateral heterostructures and more complex vdWHs consisting of more than two monolayers; \textbf{Section \ref{Photocatalyst & Photovoltaic}} provides an in-depth discussions on the  photocatalytic and photovoltaic abilities of MA$_2$Z$_4$-based SS vdWHs; \textbf{Section \ref{outlook}} provides an outlook on the challenges of MA$_2$Z$_4$-based heterostructures before the conclusion of this review. 

\section{\label{sec:vdWH}Overview of MA\textsubscript{2}Z\textsubscript{4}-based Heterostrutures}


The $\mathrm{MA_{2}Z_{4}}$-based heterostructures can be classified, based on the metallic or semiconducting nature of the constituent monolayers, into two main categories: (A) Metal/Semiconductor (MS) vdWHs; and (B) Semiconductor/Semiconductor (SS) vdWHs. For MS vdWHs, two subcategories can be defined: (A1) Ohmic contact in which charge injection can be efficiently achieved without traversing through a potential barrier; and (A2) Schottky contact in which charge injection is significantly impeded by a potential barrier arising from the mismatch of the metal work function and semiconductor band edge energies. 

For SS vdWHs, four subcategories arise depending on the relative alignment of the conduction band minimum (CBM) and the valence band maximum (VBM) of the contacting semiconductors: (B1) Type-I vdWHs in which the CBM and VBM originate exclusively from only one of the two contacting semiconductors (straddling band gap); (B2) Type-II vdWHs in which the CBM and VBM are contributed distinctively from two different semiconductors (staggered band gap); (B3) Type-III vdWHs in which the CBM of one semiconductor crosses energetically with the VBM of the other semiconductor (broken band gap), such that this overlapping leads to an overall metallic (or semimetallic) nature of the resulting heterostructure; and (B4) Type-H vdWHs in which the VBM of the heterostructure is contributed by the VBM states of the two different semiconductors (hybridized).

$\mathrm{MA_{2}Z_{4}}$-based heterostructures can be generally constructed via lateral stitching or vertical stacking configurations to form covalently bonded lateral heterostructures or vdWHs, respectively (see Fig. 2). 
Motivated by the continual discovery of new 2D materials and the advancement of vdW engineering in recent years \cite{Castellanos2014,Gurarslan2014,Cui2018,Desai2016,Zaretski2015_01,Zaretski2015_02,Bae2017}, the physical and chemical properties of $\mathrm{MA_2Z_4}$-based vdWHs have been intensively studied using DFT simulations in recent years. DFT simulations have predicted a large variety of $\mathrm{MA_2Z_4}$-based vdWHs having promising application in the areas of optoelectronics \cite{Liu2022, Cai2021_MoSe2, Cai2021_WSe2,Pham2021,Nguyen2022_C3N4,Wang2021_MoGe2N4, Guo2022_BP}, in clean energy research pertaining to photocatalysts \cite{He2022,Zeng2021,Xuefeng2022,Liu2022}, heat retention \cite{Hussain2022_thermal}, and beyond conventional charge transport devices such as spintronics \cite{Ren2022} and valleytronics \cite{Zhao2021}.
It should be noted that the Anderson rule, which determines the band alignment of heterostructures solely by aligning the vacuum levels of the constituent materials without considering the interface effects, often yield incorrect predictions of the band alignment \cite{Besse2021} due to the omission of interlayer interactions, interfacial charge transfer, interfacial dipole \cite{Tersoff1984}, and orbital hybridization \cite{Koda2017} that are inevitably present in vdWHs. Explicit direct DFT calculations are thus neccessarily in order to predict the band alignment and contact types of $\mathrm{MA_2Z_4}$-based heterostructures.

In the following sections, we systematically review the physical properties and the applicaitons of $\mathrm{MA_2Z_4}$ heterostructures based on the contact classifications outlined above. The physical properties of $\mathrm{MA_2Z_4}$-based contact hetersotructures are summarized in Tables I, II, III, IV.

\section{\label{MS_Contact}Metal/Semiconductor Heterostructures}

\subsection{$\mathrm{MA_2Z_4}$-based MS Contacts}

\begin{figure*}
\includegraphics[width=0.8\textwidth]{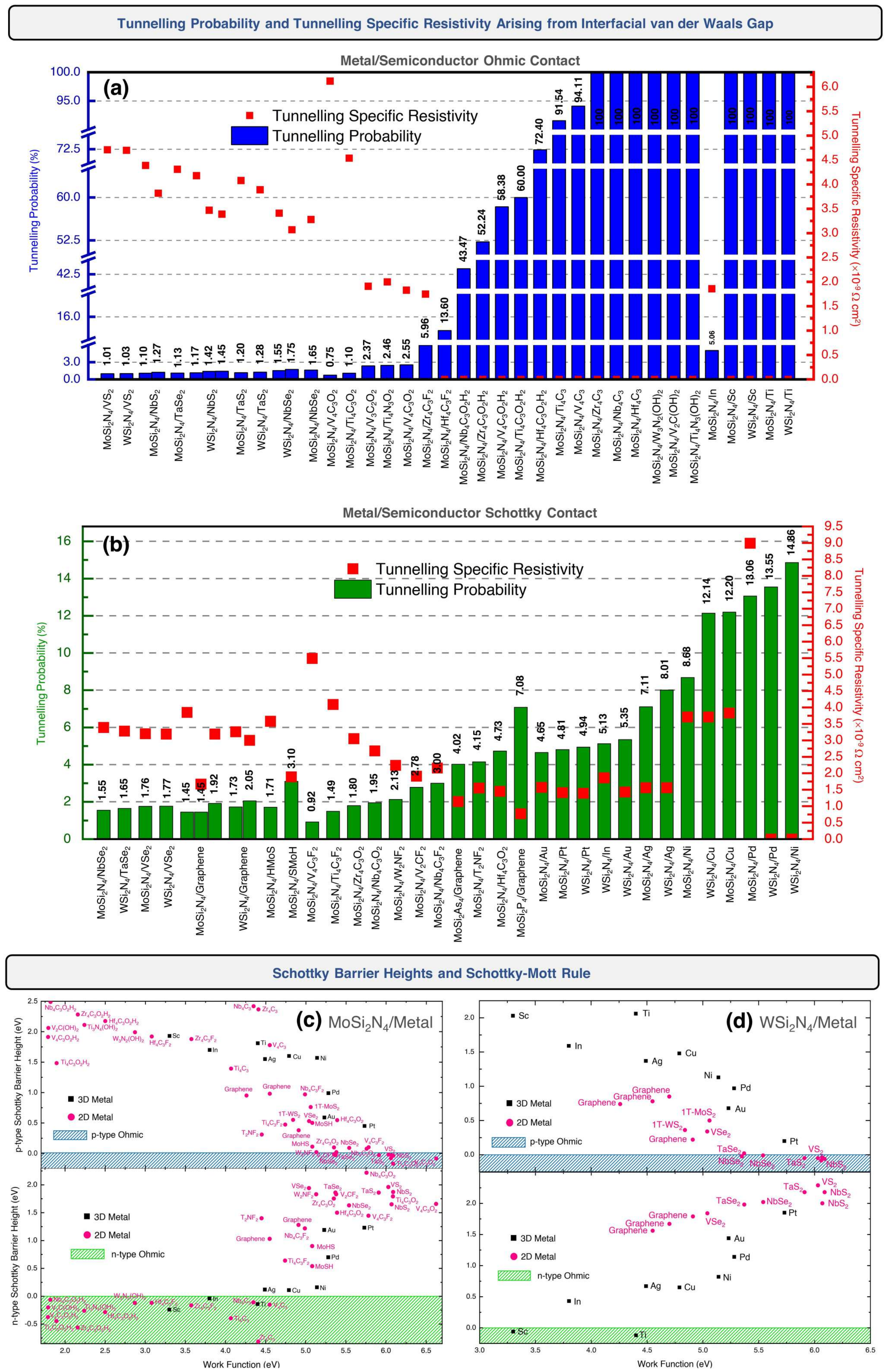}
\caption{\label{Fig3} \textbf{Electrical properties of Ohmic and Schottky contacts}. Tunneling Probability ($T$) and Tunnelling Specific Resistivity ($\rho_{t}$) of Metal/Semiconductor (MS) (a) Ohmic and  (b) Schottky contacts. Bar graph and scatter plots are used for $T$ and $\rho_{t}$. (c) $p(n)$-type Schottky barrier heights of $\mathrm{MoSi_2N_4}$ (d) $p(n)$-type Schottky barrier heights (SBH) of $\mathrm{WSi_2N_4}$. Data are reproduced from Ref. \cite{Wang2021_MA2Z4,Tho2022,Liang,Nguyen2022_MoSH,Li2023_graphene,Ma2023,He2023,Zhang2023_MXene}.}
\end{figure*}

\begin{figure*}
\includegraphics[width=0.9\textwidth]{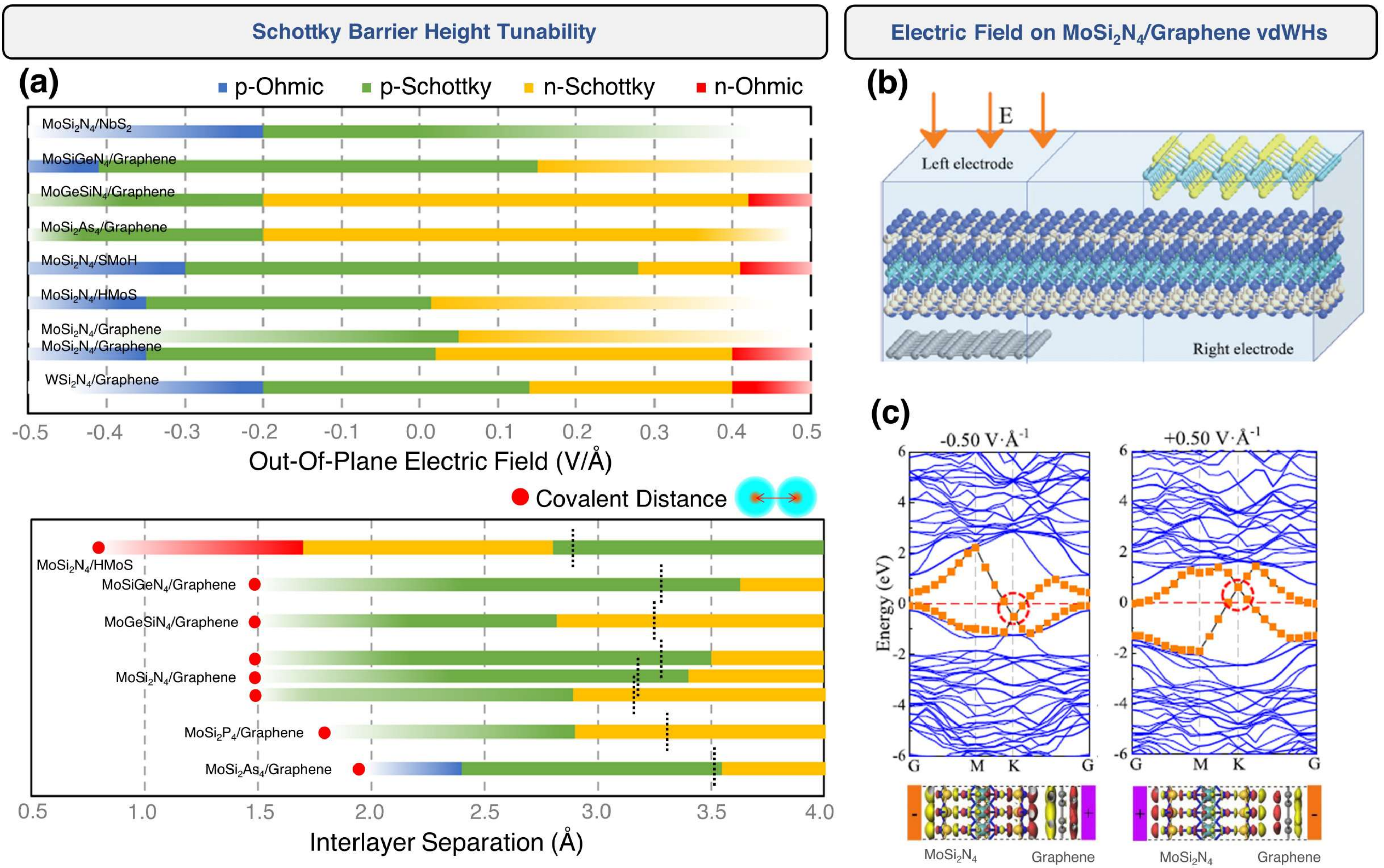}
\caption{\label{Fig4} \textbf{Tunable Ohmic/Schottky contact types and device applications}. The MS contact type can be tuned by applying (a) out-of-plane electric field (Top) and vertical strain (Bottom). 
Red circles show the covalent distances of the respectively vdWHs where the contact type beyond this distance are likely to form chemical bonds. Black dotted vertical lines denote the equilibrium distance of each vdWH. Data taken from Ref. \cite{Nguyen2022_MoSH,Li2023_graphene,Ma2023,Cao2021,Binh2021,Guo2022_Graphene,Yuan2022}. (b) Schematic of a nanodiode based on $\mathrm{graphene/MoSi_2N_4/NbS_2}$ \cite{Cao2021}. The nanodiode can be configured between lateral $p$-$p$ and $p$-$n$ junction by applying an external electric field at the $\mathrm{graphene/MoSi_2N_4}$ junction while keeping the $\mathrm{MoSi_2N_4/NbS_2}$ junction as $p$-type Schottky. (c) Under sufficiently an electric field of $\pm 0.5$ \text{V/\AA}, the Dirac point of $\mathrm{MoSi_2N_4/graphene}$ is shown to shift above (below) the Fermi level which corresponds to $p (n)$-doping of graphene. Blue (Orange) plots denote the electronic band structure contributed from $\mathrm{MoSi_2N_4(graphene)}$. Charge density distribution across the vdWH is shown at the bottom. Red (Yellow) region denotes electron accumulation (depletion) region. (b) Reproduced from Cao $et \ al.$, Appl. Phys. Lett. \textbf{118}, 013106 (2021), with the permission from AIP Publishing. (c) Reproduced with permission from Yuan $et \ al.$, ACS Appl. Electron. Mater. \textbf{4}, 2897 (2022). Copyright 2022 American Chemical Society.}
\end{figure*}

\textbf{Metal/Semiconductor Contact Physics.} At the interface of a MS Schottky contact, an energy barrier called Schottky barrier height (SBH) is formed, caused by the difference between the work function of metal and electron affinity ($n$-type Schottky barrier)/ionization potential ($p$-type Schottky barrier) of the semiconductor. 
Electric current flows across metal and semiconductor components in many modern electronics, therefore the study of SBH formation and its tunability are crucial in mitigating device performance. 
The SBH can be controlled by appropriately aligning the metal work function with the semiconductor band edge energies according to the Schottky-Mott rule:
\begin{subequations}\label{eq:SM}
\begin{equation}
    \Phi_B^{(e)} = SW_\text{M} - E_\text{CBM}
\end{equation}
\begin{equation}
    \Phi_B^{(h)} = E_\text{VBM} - SW_\text{M}
\end{equation}
\end{subequations}
where $\Phi_B^{(e/h)}$ is the electron/hole SBH, $E_\text{CBM/VBM}$ is the electron affinity/ionization potential (or the energies of CBM/VBM relative to vacuum), $W_\text{M}$ is the metal work function, and $S$ is the slope parameter which characterizes the extend of Fermi level pinning at the interface \cite{tung2014physics} with $S=1$ denoting the Schottky-Mott limit of a non-interacting MS contact.
Most 3D/2D and 3D/3D MS contacts deviate from the Schottky-Mott limit due to the formation of metal-induced gap states (MIGS) and orbital hybridization effects \cite{PhysRevX.4.031005}.
Such strong interfacial interactions arise due to the presence of many dangling bonds in 3D materials, thus when used as electrode contacts with semiconductor channels, high densities of MIGS can form which pins the Fermi level to a narrow range of values. 
Monolayer metals, on the other hand, do not have dangling bonds on their surfaces. The \textcolor{black}{absence} of MIGS in 2D/2D vdWH metal/semiconductor contact thus suppresses the Fermi level pinning effect \cite{FLP}. 
The absence of SH in MS Ohmic contacts provides a low resistance path for charge carriers transport, which is beneficial for energy-efficient device operation \cite{infomat}. Although SBH is undesirable for electrode applications (such as metal contact to a transistor), they form important functional building blocks for non-reciprocal electronic devices, such as Schottky diodes \cite{sze2021physics}. Understanding the physics of MS contact formation and the engineering of SBH are thus tremendously important for electronics and optoelectronics applications \cite{tung2014physics}.

\textcolor{black}{
For 2D/2D MS contacts, the MIGS and FLP are substantially suppressed due to the weak vdWs coupling between the two constituent monolayers \cite{FLP}. Nevertheless, substantial deviation of the SBH from the Schottky-Mott rule [Eq. (\ref{eq:SM})] can still occur due to the formation of interface dipole potential \cite{li2022revealing} and strain-induced band edge energy shifting \cite{Tho2022}. For MoSi$_2$N$_4$ and WSi$_2$N$_4$ -- key representative semiconducting monolayers of the MA$_2$Z$_4$ family \cite{Hong2020}, the MIGS and FLP effects is further suppressed due to two unique aspects: (i) The `ultrathick' morphology of MA$_2$Z$_4$ spatially separates the contacting metal further away from the semiconductor, thus minimizing the wave function penetration of the metallic states into the semiconductor; (ii) The presence of an chemically inert outer Si-N sublayer protects the VBM and CBM states residing in the inner Mo-N (or W-N) corelayer from being hybridized by the contacting metal \cite{Wang2021, wu2022prediction}, even for the case where the metal is closely contacted to MoSi$_2$N$_4$ or WSi$_2$N$_4$ monolayers \cite{Wang2021}. The combination of these two aspects allows the SBH to be more effectively tuned by appropriate choosing the contacting metal. This aspect is in stark contrast to many other 2D semiconductors such as MoS$_2$ in which the FLP effect is severe, thereby making $\mathrm{MA_2Z_4}$ monolayers attractive candidates for designing MS Ohmic or Schottky contacts.  }


\textcolor{black}{In addition to SBH, the interfacial potential barrier situated at the vdW gap separated the two contacting monolayers may impede charge ejection efficiency.}
The charge injection efficiency can be characterized by the tunnelling probability ($T$) which is obtained via the Wentzel-Kramers-Brillouin (WKB) approximation, and the tunnelling specific resistivity ($\rho_{t}$) via the Simmons formula \cite{Simmons1963}:
\begin{subequations}
\begin{equation}\label{tunneling_probability}
T(\Phi_{TB},d_{TB})=\exp\left(\frac{-2d_{TB}\sqrt{2m_{e}\Phi_{TB}}}{\hbar}\right)
\end{equation}
\begin{equation}\label{tunneling_resistivity_intermediate}
\rho_{t}(\Phi_{TB},d_{TB})=\frac{\pi^{2}\hbar d_{TB}^{2}}{e^{2}\left(\frac{\sqrt{2m_{e}\Phi_{TB}}}{\hbar}-1\right)}T^{-1}
\end{equation}
\end{subequations}
where $e$, $\hbar$, $m_{e}$ is the charge of the electron, reduced Planck constant, mass of the electron, respectively, and $\Phi_{TB}$ and $d_{TB}$ are the tunnelling barrier height and width, respectively. Equation \ref{tunneling_resistivity_intermediate} is used under intermediate bias voltages, which may give negative unphysical results of its tunnelling current when $d_{TB}$ is too small \cite{Wang2021}. The very low voltage approximation of Simmons formula \cite{Simmons1963} can be used [Equation \ref{tunneling_resistivity_low}]: 
\begin{equation}\label{tunneling_resistivity_low}
\rho_{t}(\Phi_{TB},d_{TB})=\frac{8\pi^{2}\hbar^{2} d_{TB}}{3e^{2}\sqrt{2m_{e}\Phi_{TB}}}T^{-1}
\end{equation}
In the following section, we review the SBH, $T$ and $\rho_t$ of MA$_2$Z$_4$-based MS contacts under both 3D/2D and 2D/2D configurations.

\textbf{Bulk (3D) vs 2D metals.} The calculated $\rho_{t}$ of MoSi$_2$N$_4$ and WSi$_2$N$_4$-based MS contacts with several representative 3D metals are found to be in the order of $\mathrm{10^{-9}cm^{2}}$ \cite{Wang2021}, which is comparable to low-contact-resistance Bi/MoS$_2$ \cite{Shen2021}. In Fig. \ref{Fig3}(a), we rank MoSi$_2$N$_4$ and WSi$_2$N$_4$ Ohmic contacts in increasing values of $T$, and also the corresponding $\rho_{t}$. Generally, $\rho_{t}$ decreases with increasing $T$ since large tunneling probability promotes charge transport across the contact. The $\rho_{t}$ is completely absent with $T$ = 100\% when Sc and Ti are used as 3D metal contacts, thus indicating Sc and Ti as promising candidates for achieving high charge ejection efficiency in MoSi$_2$N$_4$ and WSi$_2$N$_4$. 
The $T$ of 3D/2D MS contacts are generally higher than those of 2D/2D MS contacts due to the stronger coupling between metal and semiconductor in 3D/2D MS contacts \cite{Nguyen2022_MoSH} [see Fig. \ref{Fig3}(b)].
The SBHs of MoSi$_2$N$_4$-and WSi$_2$N$_4$-based 3D/2D and 2D/2D MS contacts are summarized in Figs. \ref{Fig3}(c) and (d). 
3D metals (In, Sc, Ti) predominantly form $n$-type Ohmic contact, whereas 2D metals ($\mathrm{NbS_2, TaSe_2, TaS_2, VS_2, NbSe_2}$) predominantly form $p$-type Ohmic contact. 

Recent computational designs of MXene metal contacts to MoSi$_2$N$_4$ reveals that $n$-type Ohmic contacts (e.g. W$_3$N$_2$(OH), V$_2$C(OH)$_2$, Ti$_4$C$_3$) can also be achieved due to the low work function nature such MXenes and the presence of an appropriate interface dipole potential that leads to a proper metal work function alignment with the conduction band continuum of MoSi$_2$N$_4$ \cite{He2023,Zhang2023_MXene}. Importantly, the presence of both $n$-and $p$-type Ohmic contacts in MXene/MoSi$_2$N$_4$ heterostructures suggest MXene metal contact can potentially enable the complementary metal-oxide-semiconductor (CMOS) implementation of MoSi$_2$N$_4$ in which both $p$-type and $n$-type devices are needed \cite{Li2023}, thus providing a route to greatly expanding the potential usefulness of MoSi$_2$N$_4$-based electronic devices.
\textcolor{black}{
It should be noted that achieving Ohmic band alignment across a 2D/2D MS contact does not immediately correspond to a low contact resistance. 
Interfacial tunnelling potential barrier, as discussed above, may still be present due to the vdW coupling nature of the MXene/MA$_2$Z$_4$ contact heterostructure. The tunnelling-specific contact resistivity of MXene/MA$_2$Z$_4$, however, has yet to be catalogued in the literature thus far, and we expect a comprehensive study on the tunneling charge transport to yield important light on the feasibility and performance of MXene/MA$_2$Z$_4$ contacts. 
}


\begin{table}[t]
{
\caption{\label{table_mobility}Mobilities $\mu_{x/y} (\times10^{3} cm^{2} V^{-1}s^{-1})$ of electrons and holes in the two in-plane orthogonal directions denoted by x and y, of vdWHs and their constituent monolayers. Superscripts $^a$ and $^z$ denote carrier mobility along the armchair and zig-zag directions, respectively, as mentioned in the papers. For Janus-MoSiGeN$_4$/Janus-MoSSe, superscripts denote carrier mobility along the path in reciprocal space as mentioned in Ref. \cite{Zhang2023}.}
\scalebox{0.8}{
\begin{tabular}{cccccc}
\hline
\hline

\multirow{2}{*}{\textbf{Materials}} & \multicolumn{2}{c}{$\mu_{x}$}& \multicolumn{2}{c}{$\mu_{y}$} & \multirow{2}{*}{\textbf{Ref.}} \\ 


& Electron & Hole &Electron&Hole &  \\ \hline

$\mathrm{MoSe_{2}}$ & $\mathrm{^a}$0.15 & $\mathrm{^a}$0.98 & $\mathrm{^z}$0.16 & $\mathrm{^z}$1.01 & \cite{Cai2021_MoSe2} \\ 
& $\mathrm{^z}$0.213510 & $\mathrm{^z}$0.562150 & $\mathrm{^a}$0.142057 & $\mathrm{^a}$0.063423 & \cite{Pei2023} \\

$\mathrm{WSe_{2}}$ & $\mathrm{^a}$9.08 & $\mathrm{^a}$9.48 & $\mathrm{^z}$15.26 & $\mathrm{^z}$9.02 & \cite{Cai2021_WSe2}\\ 
& $\mathrm{^z}$0.273305 & $\mathrm{^z}$1.794131 & $\mathrm{^a}$0.133276 & $\mathrm{^a}$0.179049 & \cite{Pei2023} \\

$\mathrm{InSe}$ & 1.61951 & 0.1349 & 1.77916 & 0.0657 & \cite{He2022}
\\ 

$\mathrm{BP}$ & $\mathrm{^z}$8.14& $\mathrm{^z}$2.50& $\mathrm{^a}$6.36& $\mathrm{^a}$2.01& \cite{Guo2022_BP} \\ 

\multirow{1}{*}{$\mathrm{MoSi_{2}N_{4}}$} & 0.27& 1.2 & 0.27 & 1.2 & \cite{Hong2020} \\ 
 & 2.32709 & 0.28510 & 2.97531 & 0.26947 & \cite{He2022} \\ 
 & 0.98376 & 0.10658 & 1.22954 & 0.09998 &  \cite{Zhao2022} \\ 
 & $\mathrm{^z}$0.37749 & $\mathrm{^z}$0.609322 & $\mathrm{^a}$0.37147 & $\mathrm{^a}$0.60984 & \cite{Yu2021} \\ 
 & $\mathrm{^z}$1.73 & $\mathrm{^z}$0.172 & $\mathrm{^a}$1.721 & $\mathrm{^a}$0.174 & \cite{Ren2023} \\
 
$\mathrm{MoSi_{2}P_{4}}$ & $\mathrm{^z}$0.47& $\mathrm{^z}$0.82& $\mathrm{^a}$0.58& $\mathrm{^a}$1.16& \cite{Guo2022_BP} \\ 

$\mathrm{MoSiGeN_{4}}$ & 4.8832 & 0.0065 & 7.0747 & 0.0065 & \cite{Lv2022} \\
& $\mathrm{^z}$0.40423 & $\mathrm{^z}$0.01047 & $\mathrm{^a}$0.3793 & $\mathrm{^a}$0.00827 & \cite{Yu2021} \\ 

$\mathrm{WSiGeN_{4}}$ & $\mathrm{^z}$0.48213 & $\mathrm{^z}$0.004 & $\mathrm{^a}$0.44066 & $\mathrm{^a}$0.0028 & \cite{Yu2021} \\ 

$\mathrm{WSi_{2}N_{4}}$ & 1.11797 & 0.10649 & 1.28499 & 0.11448 & \cite{Zhao2022} \\ 
& $\mathrm{^z}$1.732925 & $\mathrm{^z}$0.273762 & $\mathrm{^a}$1.155009 & $\mathrm{^a}$0.266595 & \cite{Pei2023} \\
& $\mathrm{^z}$2.172 & $\mathrm{^z}$0.205 & $\mathrm{^a}$2.162 & $\mathrm{^a}$0.203 & \cite{Ren2023} \\

$\mathrm{CrSi_{2}N_{4}}$ & $\mathrm{^z}$0.826 & $\mathrm{^z}$0.085 & $\mathrm{^a}$0.125 & $\mathrm{^a}$0.075 & \cite{Ren2023} \\

$\mathrm{HfSi_{2}N_{4}}$ & $\mathrm{^z}$0.206 & $\mathrm{^z}$1.182 & $\mathrm{^a}$0.966 & $\mathrm{^a}$0.706 & \cite{Ren2023} \\

$\mathrm{MoGe_{2}N_{4}}$ & $\mathrm{^z}$1.226 & $\mathrm{^z}$1.248 & $\mathrm{^a}$1.225 & $\mathrm{^a}$1.045 & \cite{Ren2023} \\

$\mathrm{MoSi_{2}As_{4}}$ & $\mathrm{^z}$0.386 & $\mathrm{^z}$1.874 & $\mathrm{^a}$0.446 & $\mathrm{^a}$1.476 & \cite{Ren2023} \\

$\mathrm{MoSi_{2}P_{4}}$ & $\mathrm{^z}$0.932 & $\mathrm{^z}$1.889 & $\mathrm{^a}$0.736 & $\mathrm{^a}$2.169 & \cite{Ren2023} \\

$\mathrm{TiSi_{2}N_{4}}$ & $\mathrm{^z}$0.141 & $\mathrm{^z}$1.695 & $\mathrm{^a}$10.370 & $\mathrm{^a}$0.815 & \cite{Ren2023} \\

$\mathrm{ZrSi_{2}N_{4}}$ & $\mathrm{^z}$0.093 & $\mathrm{^z}$0.076 & $\mathrm{^a}$0.25 & $\mathrm{^a}$0.694 & \cite{Ren2023} \\

$\mathrm{MoSi_{2}N_{4}/MoSe_{2}}$ & $\mathrm{^a}$1.84 & $\mathrm{^a}$22.35 & $\mathrm{^z}$1.86 & $\mathrm{^z}$22.39 & \cite{Cai2021_MoSe2} \\ 

$\mathrm{MoSi_{2}N_{4}/WSe_{2}}$ & $\mathrm{^a}$1.44 & $\mathrm{^a}$24.86 & $\mathrm{^z}$2.54 & $\mathrm{^z}$43.28 & \cite{Cai2021_WSe2} \\ 

$\mathrm{MoSi_{2}N_{4}/InSe}$ & 12.67688 & 0.25185 & 12.20292 & 0.23630 & \cite{He2022}\\ 

$\mathrm{MoSi_{2}P_{4}/BP}$ & $\mathrm{^z}$1.15& $\mathrm{^z}$3.17& $\mathrm{^a}$0.86& $\mathrm{^a}$2.02& \cite{Guo2022_BP}\\ 

$\mathrm{MoSi_{2}N_{4}/WSi_{2}N_{4}}$ & 2.06648 & 0.31404 & 2.03585 & 0.29029 & \cite{Zhao2022} \\ 

$\mathrm{MoSiGeN_{4}/MoSiGeN_{4}}$ & 58.5223 & 0.0722 & 48.7546 & 0.0746 & \cite{Lv2022} \\ 

$\mathrm{WSi_{2}N_{4}/MoSe_{2}}$ & 1.584591 & 25.146891 & 1.187452 & 10.496258 & \cite{Pei2023} \\

$\mathrm{WSi_{2}N_{4}/WSe_{2}}$ & 2.342283 & 32.832080 & 2.461746 & 33.032965 & \cite{Pei2023} \\

$\mathrm{MoSiGeN4/SMoSe}$&$\mathrm{^{K \rightarrow \Gamma}}$5.555902&$\mathrm{^{\Gamma \rightarrow M}}$0.38638&-&-&\cite{Zhang2023} \\

$\mathrm{MoGeSiN4/SMoSe}$&$\mathrm{^{K \rightarrow \Gamma}}$5.85706&$\mathrm{^{\Gamma \rightarrow M}}$0.24851&-&-&\cite{Zhang2023} \\

$\mathrm{MoSiGeN4/SeMoS}$&$\mathrm{^{\Gamma \rightarrow M}}$6.04843&$\mathrm{^{\Gamma \rightarrow M}}$0.26266&-&-&\cite{Zhang2023} \\

$\mathrm{MoGeSiN4/SeMoS}$&$\mathrm{^{K \rightarrow \Gamma}}$5.63101&$\mathrm{^{\Gamma \rightarrow M}}$0.22369&-&-&\cite{Zhang2023} \\

\hline
\hline
\end{tabular}
}
}
\end{table}

Beyond MoSi$_2$N$_4$ and WSi$_2$N$_4$ monolayers, CrSi$_2$N$_4$ and CrC$_2$N$_4$ from the MA$_2$Z$_4$ family \cite{Shu2023} have also been predicted computationally, using DFT and NEGF quantum transport simulations, to exhibit \emph{ideal} $n$-type Ohmic contacts with Ag and Ti bulk metals at both the vertical MS contact and the lateral contact/channel interface, which are beneficial for achieving high charge injection efficiency across the MS contacts. Figures \ref{Fig3}(c) and (d) show the SBH of $\mathrm{MA_2Z_4}$ MS contacts as a function of the metal work function ($\mathrm{W_M}$) compiled from the DFT simulation data extracted from previous literature. In general, the two extreme ends of very large and of very small $\mathrm{W_M}$ tends to form $p$-type and $n$-type Ohmic contact (or quasi-Ohmic contact with ultralow SBH), respectively, with MoSi$_2$N$_4$ and WSi$_2$N$_4$, due to the alignment of the metal fermi level with the band edges of the semiconducting monolayer. 
Particularly, 2D Mxenes exhibit a wider range of SBH as compared to bulk metals since the large variety of MXenes broader $W_M$ values \cite{He2023, Zhang2023_MXene}. 
In fact, for the 2D metals that have been investigated so far, MXenes are the only electrode materials predicted to form $n$-type Ohmic contacts with MoSi$_2$N$_4$.
As $p$-type and $n$-type contacts are both needed in the implementation of complementary metal-oxide-semiconductor (CMOS) technology which requires both $p$-and $n$-type transistors \cite{wang2022road}, MXene represents a promising 2D metal family in enabling CMOS functionality in the MA$_2$Z$_4$ family.

We further note that apart being an indispensable structure nearly present in all electronics and optoelectronics devices, MS contacts can also be used as a functional energy conversion device. For example, MoN$_2$/MoSi$_2$N$_4$ can be used as a nanogenerator when MoN$_2$ is dragged along MoSi$_2$N$_4$ in the in-plane direction \cite{Zhong2021}. An alternating voltage will be induced by the effect of sliding ferroelectricity, which is approximately eight times larger than that of the voltage supplied by a previously studied C/BN nanogenerator.

\begin{figure*}
\includegraphics[width=1\textwidth]{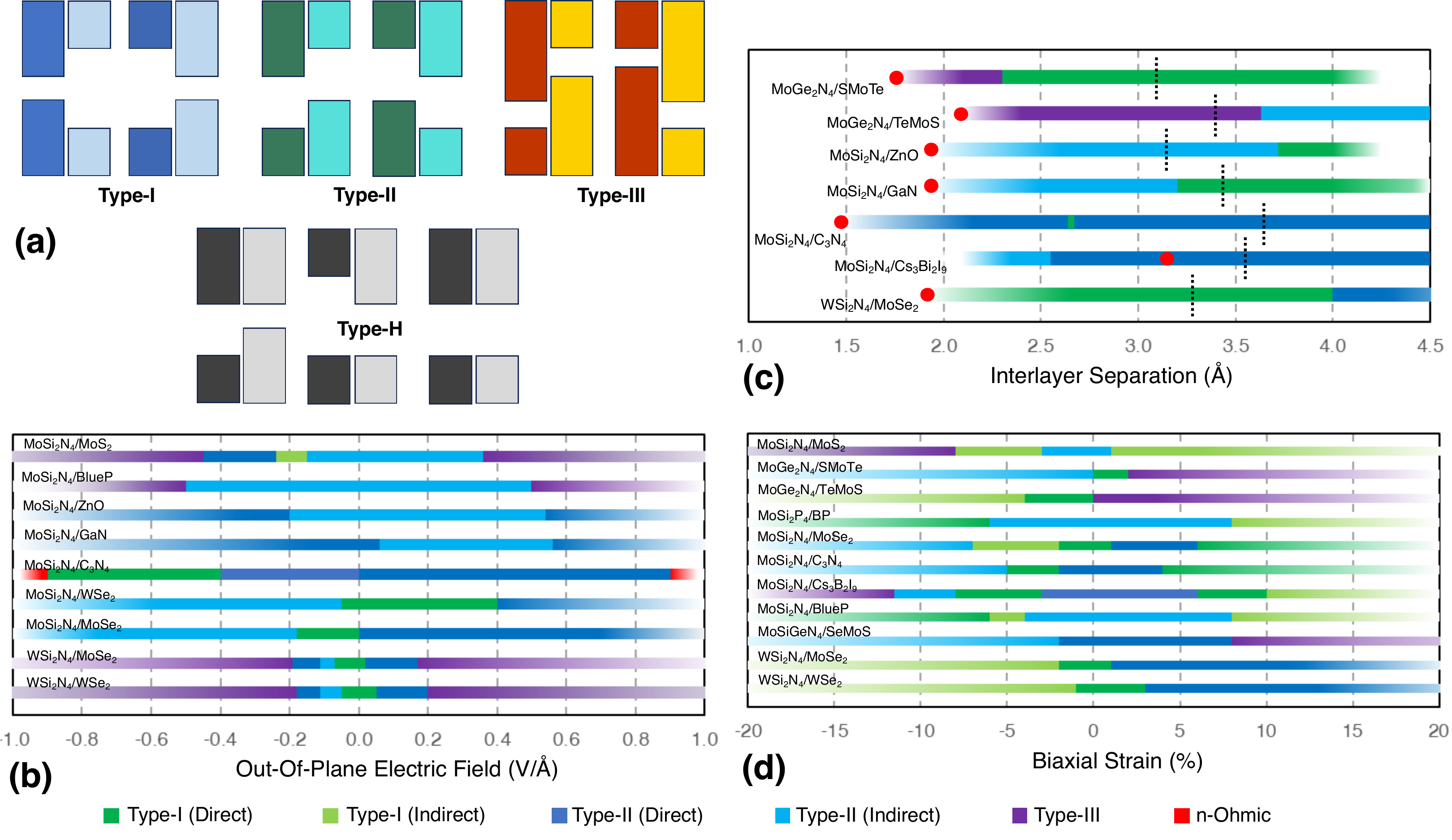}
\caption{\label{Fig5}\textbf{Tunable band alignment types of SS vdWHs}. (a) Schematic illustrating the possible band alignment category of SS vdWHs. The tuning can be achieved via (b) out-of-plane electric field, (c) out-of-plane strain and (d) biaxial strain. Colour fading at the edges and markers are similarly illustrated as in Fig. \ref{Fig4}(a). Data are taken from Ref. \cite{Guo2022_BP,Liu2022,Nguyen2022_C3N4,JinQuan2022,Cai2021_MoSe2, Cai2021_WSe2,Wang2021_MoGe2N4,Pei2023,Zhang2023,Xu2023, Fang2022}.} 
\end{figure*}

\textbf{SBH tuning through electric field, strain and Janus phase engineering.} A tunable SBH is highly desirable for electronics device applications. Akin to many 2D/2D MS contacts, the MA$_2$Z$_4$-based MS contacts exhibit electric field and strain tunable SBHs. 
In Fig. \ref{Fig4}(a)(Top), we summarize the out-of-plane electric field-tunable SBH of various reported MA$_2$Z$_4$-based MS contacts, and their transitions between Ohmic and Schottky contacts. We define the positive direction of the electric field as pointing from $\mathrm{MA_2Z_4}$ to the contacting metal. We note that $\mathrm{MoSi_2N_4/NbS_2}$ forms Ohmic contact as reported in Ref. \cite{Wang2021} and Ref. \cite{Liang}, but is reported as a $p$-type Schottky contact with ultra-low $p$-type SBH in Ref. \cite{Cao2021}. 
Such difference arises from different strain and stacking orientation which modifies the electronic bands. 
Nevertheless, field-tunable SBH is observed in a large variety of MA$_2$Z$_4$-based MS contacts. For example, over the field range of -0.5 to 0.5 V/Å, $\mathrm{MoSi_2N_4/NbS_2}$ changes its polarity from $p$-type Ohmic contact to $p$-type Schottky contact. 
For other $\mathrm{MA_2Z_4}$ MS contacts, a larger positive external electric field can substantially lower the CBM of the semiconducting monolayer to achieve a $n$-type Schottky contact or even $n$-type Ohmic contact. Particularly, $\mathrm{MA_2Z_4}$/graphene MS contacts exhibit the most versatile field-tunability which transits between various $p/n$-type Schottky and Ohmic contacts, thus rendering them useful as reconfigurable diode applications [see Fig. \ref{Fig4}(b),(c)] \cite{Cao2021,Yuan2022,Binh2021,Guo2022_Graphene}. 

Janus phase engineering provides additional freedom for tuning the SBH heights or even for changing the contact type by reversing the side of the contacting surface. Janus-MoSH when in contact with $\mathrm{MoSi_2N_4}$ can form a $p$-type Schottky barrier height of either 0.50 eV or 0.11 eV (ultra-low) by simply reversing the contacting surface of Janus-MoSH \cite{Nguyen2022_MoSH}, whereas Janus-$\mathrm{MoSiGeN_4}$/graphene contact exhibits either an $n$-type or $p$-type Schottky contact depending on its contacting surface \cite{Guo2022_Graphene}.

External strain can be used to tune the electronic structures of contact heterostructures \cite{nguyen2020interfacial, huang2019strain}. 
Vertical strain, which modifies mainly the interlayer spacing between the constituent two sub-monolayers, has been widely explored as an efficient tool to tune the SBH and contact types of $\mathrm{MA_2Z_4}$-based MS Ohmic contacts. Vertical strain can be experimentally achieved through nanomechanical engineering \cite{Dienwiebel2004}, thermal annealing \cite{Tongay2014}, or the insertion of a dielectric between the layers \cite{Fang2014}. 
Figure \ref{Fig4}(a)(Bottom) shows the transitions between Ohmic and Schottky contact of 2D/2D MS contacts that are tunable within the vertical compressive/tensile strain range. $\mathrm{MoSi_2N_4/NbS_2}$ and $\mathrm{MoSi_2N_4/SMoH}$ are only weakly tunable with vertical strain and retain $p$-type Schottky contact over a large range of vertical strain. 
$\mathrm{MoSi_2N_4/HMoS}$, $\mathrm{MoSi_2N_4/graphene}$ and Janus-MoSiGeN$_4$/graphene, on the other hand, can be tuned between Schottky and Ohmic contacts by varying their interlayer spacings \cite{Nguyen2022_MoSH, Yuan2022,Guo2022_Graphene,Binh2021}. 
It should be noted that the interlayer spacings cannot be indefinitely decreased since the contacting atoms could eventually reach the covalent distance to form strongly coupled interfaces. The covalent distance is marked by a red filled circle in Fig. \ref{Fig4}(a)(Bottom) which denotes the lower limit of vertical strain tuning. 

\begin{figure*}
\includegraphics[width=0.75\textwidth]{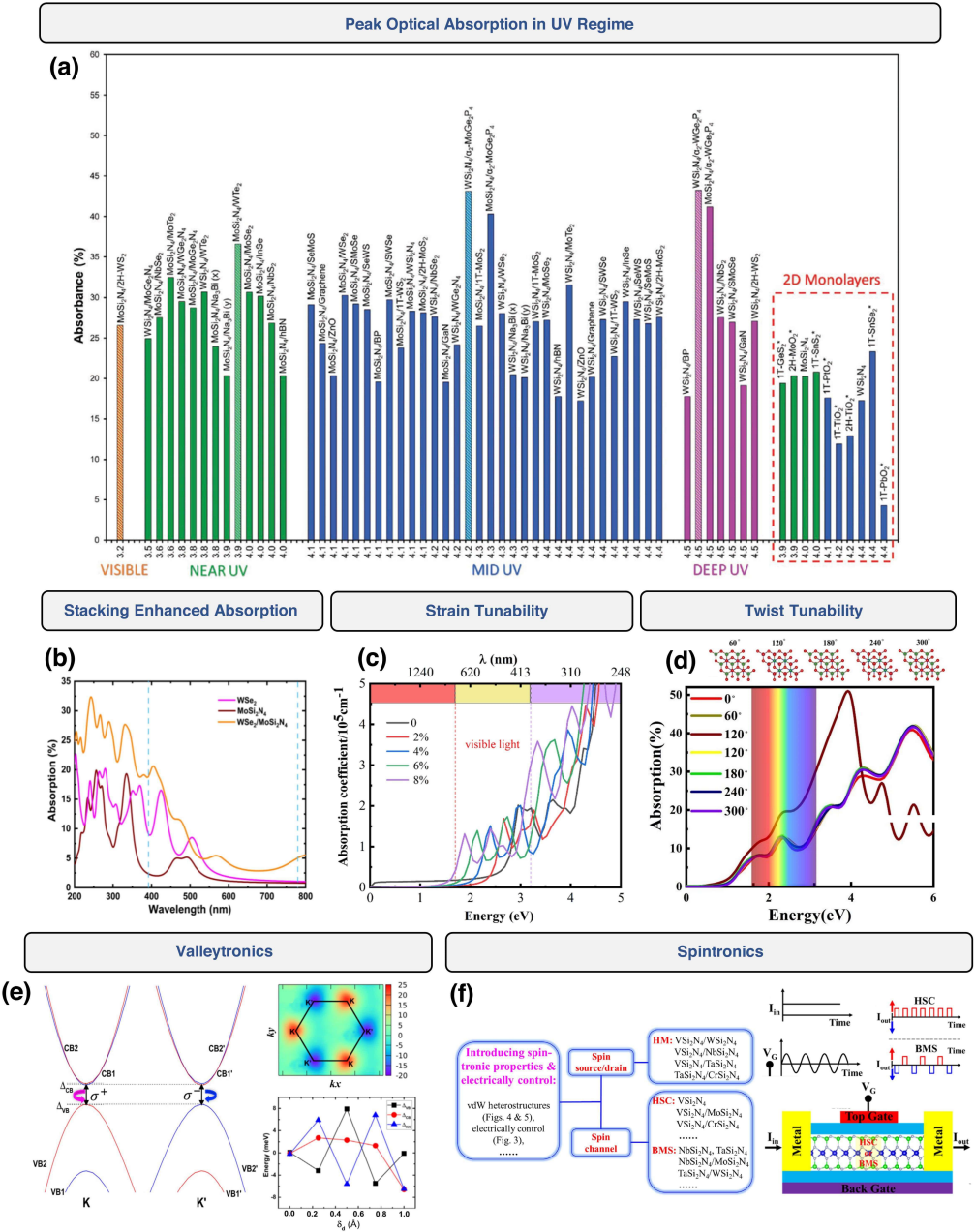}

\caption{\label{Fig6} \textbf{Optical and valleytronic properties, and spintronic device applications of MA$_2$Z$_4$ vertical SS contact heterostructures.} (a) Peak absorptions of various MA$_2$Z$_4$ vdWHs in the UV regime. (b) Optical absorption spectrum of $\mathrm{MoSi_2N_4/WSe_2}$. (c) Optical absorption spectra of $\mathrm{MoSi_2N_4/BlueP}$ under compressive biaxial strain. (d) Optical absorption curve of $\mathrm{MoSi_2P_4/BP}$ with different stacking angles. (e) Spin and valley splitting at the $K$ and $K'$ points of $\mathrm{MoSi_2N_4/CrCl_3}$. $\sigma^{+}$ and $\sigma^{-}$ denotes right/left hand circularly-polarised photon that selectively excites the electrons with different spin component at the different valleys. Shown at the right are the Berry curvature distribution of the valence band in the 1st Brillouin zone (color map), and the degree of valence and conduction valley splitting with respect to interlayer distance (line plot). (f) Schematic of a field-effect transistor (FET) spin filter or spin valve based on $\mathrm{MA_2Z_4}$ heterostructures. (a) Reproduced with permission from Tho $et \ al.$, Adv. Mater. Interfaces. \textbf{10}, 2201856 (2023). Copyright 2023 Author(s); licensed under a Creative Commons Attribution (CC BY) license. (b) Reproduced with permission from Zhang $et \ al.$, Phys. E: Low-Dimens. Syst. Nanostructures. \textbf{144}, 115429 (2022), Copyright 2022 Elsevier B.V.. (c) Reproduced with permission from Chen $et \ al.$, J. Phys. D: Appl. Phys. \textbf{55}, 215502 (2022), Copyright 2022 IOP Publishing Ltd. (d) Reproduced with permission from Guo $et \ al.$, J. Phys. Chem. C \textbf{126}, 4677 (2022), Copyright 2022 American Chemical Society. (e) Reproduced from Zhao $et \ al.$, Appl. Phys. Lett. \textbf{119}, 213101 (2021), with the permission from AIP Publishing. (f) Reproduced with permission from Ren $et \ al.$, Phys. Rev. Materials \textbf{6}, 064006 (2022), Copyright 2022 American Physical Society.}
\end{figure*}

\section{\label{SS Contact}Semiconductor/semiconductor Heterostructures}

\subsection{\label{Mobility}$\mathrm{MA_2Z_4}$-based Heterostructure Mobility}

\textbf{Carrier mobility enhancement.} The mobility of electrons and holes in 2D materials can be significantly enhanced when in heterostructures which stems from either an increased in the in-plane stiffness or a reduction of the effective mass or the deformation potential. Based on the Bardeen-Shockley deformation potential theory (BS-DPT) \cite{PhysRev.80.72, PhysRevB.94.235306}, the carrier mobility due to longitudinal acoustic (LA) phonon scattering in 2D semiconductors can be expressed as:

%
\begin{equation}\label{mobility}
\mu_{2D}=\frac{e C_{2D} \hbar^{3}}{k_{B}Tm^{*}m_{d}(E_{1})^{2}}
\end{equation}
where $C_{2D}$, $k_{B}$, $T$, $m^{*}$, $m_{d}$, $E_{1}$ is the elastic modulus, Boltzmann constant, temperature (taken as about 300 K), effective mass of the carrier in the transport direction, average effective mass of the carrier, deformation potential respectively. High carrier mobility is generally desirable for the construction of transistors \cite{Li2023_graphene}, photocatalysts \cite{Mir2020}, excitonic solar cells \cite{Gregg2003}, and many other applications that involve charge current. In transistors application, high carrier mobility directly translates to faster switching speeds and larger current-carrying capabilities, which allows  for high speed information processing and greater power output respectively. In solar cell and photocatalytic applications, higher carrier mobility increases the drift and diffusion photocurrent to the collection terminals (in solar cells) or to the oxidation/reduction sites (in photocatalysts) before carrier recombination can take place. Carrier mobility is thus an indispensable quantity for consideration when constructing modern electronics and optoelectronics devices.

Carrier mobility enhancement via vdWH engineering have been investigated for various MA$_2$Z$_4$-based SS contacts, where we summarized the result in Table \ref{table_mobility} for the electron and hole mobility of MA$_2$Z$_4$-based SS vdWHs. The carrier mobility of the constituent monolayers are also shown for a comparison. 
Generally, the carrier mobility of MA$_2$Z$_4$-based SS vdWHs are enhanced when compared to their monolayer counterparts.
For instance, $\mathrm{MoSi_2N_4/WSe_2}$ \cite{Cai2021_WSe2} and $\mathrm{MoSi_2N_4/MoSe_2}$ \cite{Cai2021_MoSe2} forms Type-I SS vdWH with a significantly larger hole mobility when compared to both the standalone MoSi$_2$N$_4$ and MoSe$_2$. The electron mobility of only $10^2$ $\mathrm{cm^{2}V^{-1}s^{-1}}$ in $\mathrm{MoSe_2}$ and in MoSi$_2$N$_4$ can be enhanced by an order of magnitude when forming $\mathrm{MoSi_2N_4/MoSe_2}$. vdWH engineering thus is an effective method for enhancing carrier mobility.  


\begin{figure*}
\includegraphics[width=1\textwidth]{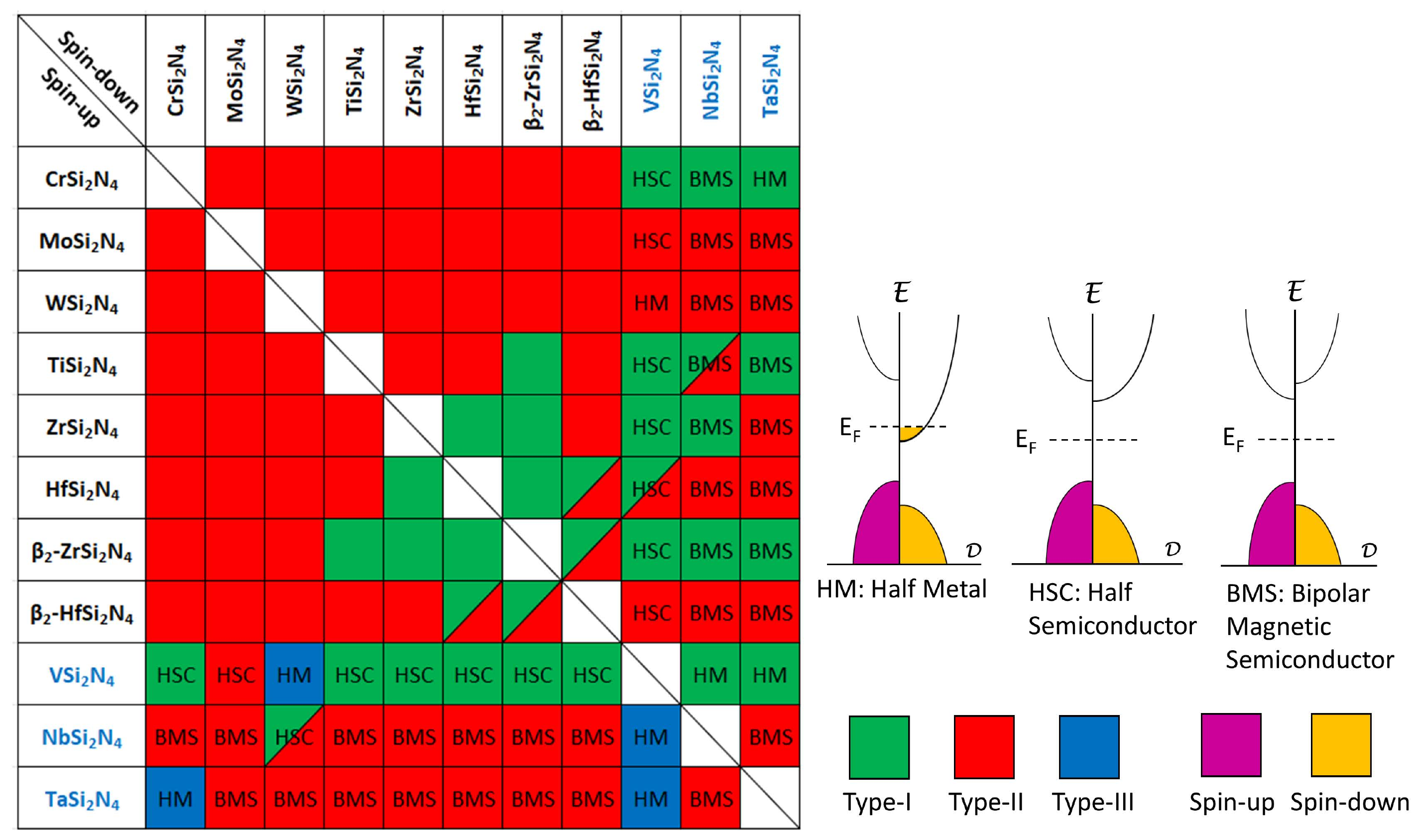}
\caption{\label{Fig7} \textbf{MA$_2$Z$_4$-based vdWHs constructed from monolayers with M = Group IV-B, V-B, VI-B element.} The band alignment is determined using Anderson's rule (except for $\mathrm{MoSi_2N_4/WSi_2N_4}$, $\mathrm{MoSi_2N_4/VSi_2N_4}$, $\mathrm{WSi_2N_4/VSi_2N_4}$ calculated using DFT). vdWHs consisting of Group V-B element exhibits novel semiconducting (half semiconductor, bipolar magnetic semiconductors) or metallic properties (half metals). Reproduced with permission from Ren $et \ al.$, Phys. Rev. Materials \textbf{6}, 064006 (2022), Copyright 2022 American Physical Society.}
\end{figure*} 



\subsection{\label{SS_Band_Alignment}$\mathrm{MA_2Z_4}$-based SS Contacts}

\textbf{Band alignment.} SS vdWHs can be classified into four band alignment types according to the relative position of the building materials' CBM and VBM. [see Fig. \ref{Fig5}(a)]: (i) Type-I, (ii) Type-II, (iii) Type-III, and (iv) Type-H. Type-I materials with direct band gap are useful in light-emitting devices such as diodes and lasers, since the photogenerated carriers easily recombine at the same monolayer. Type-II materials are useful as photocatalysts and solar cells since the photogenerated carriers of opposite polarity can be efficiently separated across different monolayers. Type-III materials are useful for engineering tunneling FET since the electrons from the monolayer with higher energy valence band can occupy the lower energy conduction band of the other monolayer through various tunneling mechanisms. Type-H materials however, are intermediate between Type-I and Type-II, which limits its potential in optoelectronics and photocatalytic applications due to not reaping the full benefits of their distinguishable properties. Type-H materials arised due to hybridization of the out-of-plane $p_z$ orbitals in the vdW gap, and are observed in $\mathrm{WSi_2N_4/WSSe}$, $\mathrm{WSi_2N_4/MoS_2}$, $\mathrm{MoSi_2N_4/SMoSe}$ \cite{Tho2022} and in vdWHs composed of TMDC monolayers \cite{Besse2021, Silveira2022}.        

\textbf{Enhanced optical absorption.} Figures \ref{Fig5}(b)-(d) shows the band alignment tunability of SS contact vdWHs through the application of out-of-plane electric field and strain (in-plane and vertical). We define the direction of the positive electric field as pointing from $\mathrm{MA_2Z_4}$ to the other semiconductor. The in-plane strain being applied is the biaxial strain ($\epsilon$) defined as $\epsilon=(a-a_{0})/a_{0}\times100$, where $a$ and $a_{0}$ are the strained and unstrained lattice constant respectively. 2D materials are able to withstand a larger strain as compared to 3D materials due to their delayed plastic deformation and fracture phenomena \cite{Tian2021}. In the $\mathrm{MA_2Z_4}$ family, $\mathrm{MoSi_2N_4}$ has a tensile strength of 50.6 GPa and an in-plane critical strain of 19.5\%, making it a suitable material to build vdWHs that show resistance to bond breakage when strain is applied to alter their physical properties. The bandgaps and band edge positions of the SS vdWHs are much more sensitive to biaxial strain than vertical strain [see Figs. \ref{Fig5}(c),(d)], rendering the first strain approach useful for constructing optoelectronics that operate in different light absorption regimes. In general, the $\mathrm{MA_2Z_4}$-based SS contacts show better optical absorption property than their constituent monolayer $\mathrm{MA_2Z_4}$ with absorption peaks at the ultraviolet (UV) regime [see Fig.~\ref{Fig6}(a),(b)], and the application of strain to manipulate the band alignment can be used to improve their optical absorption spectrum in this regime [see Fig.~\ref{Fig6}(c) for $\mathrm{MoSi_2N_4/BlueP}$ as a representative material] such as in the case of $\mathrm{WSi_2N_4/MoSe_2}$ \cite{Pei2023}, $\mathrm{WSi_2N_4/WSe_2}$ \cite{Pei2023}, $\mathrm{MoSi_2N_4/MoSe_2}$ \cite{Cai2021_MoSe2}, $\mathrm{MoSi_2N_4/BlueP}$ \cite{Xuefeng2022}, Janus-$\mathrm{MoGe_2N_4/MoSTe}$ \cite{Wang2021_MoGe2N4}. Optical absorption can also be varied based on the stacking orientation of the monolayers, such as in the case of $\mathrm{MoSi_2P_4/BP}$ \cite{Guo2022_BP} where five different orientations are investigated [see Fig.~\ref{Fig6}(d)].

\textbf{Valleytronics and spintronics.} Magnetic semiconducting monolayers can be stacked with other monolayers to form novel vdWHs that have potential applications in valleytronics and spintronics \cite{val1}. The broken inversion symmetry of $\mathrm{MoSi_2N_4}$ can be exploited when stacked with ferromagnetic $\mathrm{CrCl_3}$ to lift the valley degeneracy at the K and K’ valleys through the valley Zeeman effect \cite{Zhao2021}. Vertical strain can tune the amount of valley splitting thereby demonstrating the potential use of $\mathrm{MoSi_2N_4}$-based magnetic vdWHs for valleytronics devices [see Fig. \ref{Fig6}(e)]. Group V-B monolayers $\mathrm{VSi_2N_4}$, $\mathrm{NbSi_2N_4}$ and $\mathrm{TaSi_2N_4}$ are magnetic semiconductors due to one unpaired electron in their $\mathrm{dz_2}$ orbital \cite{Ren2022}. vdWHs made from these Group V-B monolayers can be stacked with other $\mathrm{MA_2Z_4}$ monolayers to form half-metals ($\mathrm{VSi_2N_4/WSi_2N_4}$, $\mathrm{VSi_2N_4/NbSi_2N_4}$, $\mathrm{VSi_2N_4/TaSi_2N_4}$, $\mathrm{TaSi_2N_4/CrSi_2N_4}$), with one spin channel exhibiting Type-III band alignment and the other spin channel having a sizable energy bandgap. $\mathrm{VSi_2N_4/MoSi_2N_4}$ on the other hand is a half-semiconductor where the VBM and CBM are contributed from the same spin channel. The realization of half-metals and half-semiconductors through stacking monolayers without the use of external perturbations or chemical modifications, offers a new alternative paradigm in the construction of spintronic devices [see Fig. \ref{Fig6}(f)]. The band alignments of various vdWHs based on M = Group IV-B, V-B, VI-B reveals that most of them are potential materials for photodetection (i.e. Type-II band alignment) and for spintronics applications (i.e. spin-polarised metals and semiconductors) [Fig. \ref{Fig7}].

\textbf{Solar-to-hydrogen (STH) efficiency.} $\mathrm{H_2}$ gas production is a solution towards clean and renewable energy. The solar-to-hydrogen (STH) efficiency which determines the efficiency of hydrogen production from photocatalytic water splitting, is required to be above 10\% for the material to be commercially viable \cite{Fu2018_intrinsic_efield}. The band edge alignment of $\mathrm{MoSi_2N_4/blue phosphorus}$ (BlueP) results in a high STH efficiency of 21.1\%, demonstrating an enhancement over that of monolayer $\mathrm{MoSi_2N_4}$ (6.5\%) and BlueP (5.1\%) \cite{Xuefeng2022}. On the other hand, the STH efficiency of $\mathrm{MoSi_2N_4/WSi_2N_4}$ which is found to be 9.1\%, falls short of the 10\% benchmark but is still higher than that of its constituent monolayers \cite{Zhao2022}. These results show that constructing vdWHs is a suitable approach to improve the photocatalytic capability of the monolayers. In Section \ref{Photocatalyst & Photovoltaic}, we discuss in details the photocatalytic properties of MA$_2$Z$_4$ monolayers and SS vdWHs in which the water-splitting performance can be externally tuned via strain and using different pH conditions.



\begin{table*}
\caption{\label{MS_2D_table}Key physical quantities of 2D/2D MS contacts. HMoS and SMoH represents the H and S surface in contact with $\mathrm{MoSi_2N_4}$, respectively. For quantities not reported, the symbol `-' is used.}

\centering
 \resizebox{\textwidth}{!}
{
\begin{tabular}{>{\centering\arraybackslash}m{1.5cm}>{\centering\arraybackslash}m{1cm}>{\centering\arraybackslash}m{2cm}>{\centering\arraybackslash}m{2cm}>{\centering\arraybackslash}m{2cm}>{\centering\arraybackslash}m{1.5cm}>{\centering\arraybackslash}m{2cm}>{\centering\arraybackslash}m{2cm}>{\centering\arraybackslash}m{2cm}>{\centering\arraybackslash}m{2cm}>{\centering\arraybackslash}m{2cm}>
{\centering\arraybackslash}m{1.5cm}}

\hline
\hline

 $\mathrm{\bf{MA_{2}Z_{4}}}$ & \bf{Metal} & \bf{Contact Type} &  \multicolumn{2}{c}{\bf{Schottky Barrier Height (eV)}} &\bf{Tunnelling Barrier Height (eV)} & \bf{Tunnelling Barrier Width (Å)} & \bf{Tunnelling Barrier Probability} & \bf{Tunnelling Specific Resistivity (10$^{-9}$ 
  $\Omega$ cm$^{2}$)} & \bf{Electric Field Tunable?} & \bf{Vertical Strain Tunable?} & \bf{Ref.} \\
  

&&&n-type&p-type&&&&&&& \\
 \hline
 
\multirow{43}{*}{$\mathrm{MoSi_{2}N_{4}}$}
  & $\mathrm{VS_2}$&Ohmic&1.96&-0.02&5.29&1.95&1.01&4.71&-&-&\cite{Liang} \\
  & $\mathrm{VSe_2}$&Schottky&1.94 &0.53&4.40&1.88&1.76&3.20&-&-&\cite{Liang} \\
& $\mathrm{NbS_2}$&Ohmic & 1.87&-0.04&5.21&1.93&1.10&4.39&-&-&\cite{Liang}\\
&&Ohmic & 1.65&-0.03&5.14&1.88&1.27&3.82&-&-&\cite{Wang2021}\\
&&Schottky&&0.042&-&-&-&-&\ding{55}&\ding{55}& \cite{Cao2021}\\
 & $\mathrm{NbSe_2}$&Schottky&1.63 &0.09&4.74&1.87&1.55&3.39&-&-&\cite{Liang}\\ 
 & $\mathrm{TaS_2}$&Ohmic&1.86&-0.03&5.12&1.91&1.2&4.08&-&-&\cite{Liang}\\
 & $\mathrm{TaSe_2}$&Ohmic &1.86& -0.03&5.14&1.93&1.13&4.31&-&-&\cite{Liang}\\
 & Graphene &Schottky&1.03&0.98&4.04&1.92&1.92&3.19&-&-&\cite{Liang} \\
 &&Schottky&1.28&0.38&4.45&1.96 &1.45&3.85&-&-&\cite{Wang2021} \\
&&Schottky&1.14&1.06&-&-&-&-&\checkmark&\checkmark&\cite{Yuan2022}\\
 &&Schottky&0.92&0.80&-&-&-&-&\checkmark&\checkmark&\cite{Cao2021}\\
 &&Schottky&0.68&0.97&9.98&1.3&1.45&1.67&\checkmark&\checkmark& \cite{Li2023_graphene} \\
&HMoS&Schottky&0.54&0.5&4.04&1.98&1.71&3.58&\checkmark&\checkmark&\cite{Nguyen2022_MoSH} \\
 & SMoH &Schottky&0.9&0.11&4.27&1.63&3.10&1.89&\checkmark&\checkmark&\cite{Nguyen2022_MoSH} \\
 & Ti$_4$C$_3$ &Ohmic&-0.396&1.393&0.124&0.245&91.54&1.6$\times10^{-2}$&-&-&\cite{He2023}\\
 & Ti$_4$C$_3$F$_2$ &Schottky&0.639&0.473&4.057&2.037&1.49&4.09&-&-&\cite{He2023}\\
 & Ti$_4$C$_3$O$_2$ &Ohmic&1.791&-0.17&5.002&1.967&1.1&4.54&-&-&\cite{He2023}\\
 & Ti$_4$C$_3$O$_2$H$_2$ &Ohmic&-0.446&1.483&0.684&0.603&60&2.57$\times10^{-2}$&-&-&\cite{He2023}\\
 & V$_4$C$_3$ &Ohmic&-0.151&1.779&0.106&0.182&94.11&1.25$\times10^{-2}$&-&-&\cite{He2023}\\
 & V$_4$C$_3$F$_2$ &Schottky&1.445&0.1&5.015&2.044&0.92&5.49&-&-&\cite{He2023}\\
 & V$_4$C$_3$O$_2$ &Ohmic&1.657&-0.085&5.573&2.023&0.75&6.12&-&-&\cite{He2023}\\
 & V$_4$C$_3$O$_2$H$_2$ &Ohmic&-0.375&1.913&0.637&0.68&58.38&2.79$\times10^{-2}$&-&-&\cite{He2023}\\
 & Zr$_4$C$_3$ &Ohmic&-0.813&2.366&0&0&100&0&-&-&\cite{He2023}\\
 & Zr$_4$C$_3$F$_2$ &Ohmic&-0.164&1.879&2.866&1.626&5.96&1.75&-&-&\cite{He2023}\\
 & Zr$_4$C$_3$O$_2$ &Schottky&1.753&0.098&4.511&1.847&1.8&3.05&-&-&\cite{He2023}\\
 & Zr$_4$C$_3$O$_2$H$_2$ &Ohmic&-0.561&2.281&0.716&0.749&52.24&3.58$\times10^{-2}$&-&-&\cite{He2023}\\
 & Nb$_4$C$_3$ &Ohmic&-0.109&2.418&0&0&100&0&-&-&\cite{He2023}\\
 & Nb$_4$C$_3$F$_2$ &Schottky&1.217&0.97&3.901&1.733&3&2.16&-&-&\cite{He2023}\\
 & Nb$_4$C$_3$O$_2$ &Schottky&2.217&0.075&4.723&1.768&1.95&2.68&-&-&\cite{He2023}\\
 & Nb$_4$C$_3$O$_2$H$_2$ &Ohmic&-0.065&2.489&0.902&0.856&43.47&4.38$\times10^{-2}$&-&-&\cite{He2023}\\
 & Hf$_4$C$_3$ &Ohmic&-0.915&2.711&0&0&100&0&-&-&\cite{He2023}\\
 & Hf$_4$C$_3$F$_2$ &Ohmic&-0.118&1.921&2.304&1.283&13.6&1.31$\times10^{-1}$&-&-&\cite{He2023}\\
 & Hf$_4$C$_3$O$_2$ &Schottky&1.5&0.5453&3.981&1.493&4.73&1.45&-&-&\cite{He2023}\\
 & Hf$_4$C$_3$O$_2$H$_2$ &Ohmic&-0.284&2.174&0.419&0.487&72.4&2.19$\times10^{-2}$&-&-&\cite{He2023}\\
 &T$_2$NF$_2$&Schottky&1.4&0.31&4.12&1.53&4.15&1.55&-&-&\cite{Zhang2023_MXene} \\
 &W$_2$NF$_2$&Schottky&1.83&0.02&5.18&1.65&2.13&2.24&-&-&\cite{Zhang2023_MXene} \\
 &V$_2$CF$_2$&Schottky&1.83&0.02&4.72&1.61&2.78&1.91&-&-&\cite{Zhang2023_MXene} \\
 &V$_3$C$_2$O$_2$&Ohmic&1.92&-0.2&5.48&1.56&2.37&1.91&-&-&\cite{Zhang2023_MXene} \\
 &V$_4$C$_3$O$_2$&Ohmic&1.79&-0.1&5.34&1.55&2.55&1.83&-&-&\cite{Zhang2023_MXene} \\
 &Ti$_4$N$_3$O$_2$&Ohmic&1.76&-0.03&5.04&1.61&2.46&2&-&-&\cite{Zhang2023_MXene} \\
 &W$_3$N$_2$(OH)$_2$&Ohmic&-0.12&1.99&0&0&100&0&-&-&\cite{Zhang2023_MXene} \\
 &V$_2$C(OH)$_2$&Ohmic&-0.2&2.06&0&0&100&0&-&-&\cite{Zhang2023_MXene} \\
 &Ti$_4$N$_3$(OH)$_2$&Ohmic&-0.26&2.11&0&0&100&0&-&-&\cite{Zhang2023_MXene} \\
 \hline

\multirow{10}{*}{$\mathrm{WSi_{2}N_{4}}$}
  & $\mathrm{VS_2}$&Ohmic&2.29 &-0.05&5.19&1.96&1.03&4.70&-&-&\cite{Liang} \\
  & $\mathrm{VSe_2}$&Schottky&1.84 &0.34&4.39&1.88&1.77&3.19&-&-&\cite{Liang} \\
& $\mathrm{NbS_2}$&Ohmic &2.18&-0.06&5.12&1.92&1.17&4.18&-&-&\cite{Liang}  \\
& &Ohmic &2.00&-0.04&5.04&1.84&1.45&3.39&-&-&\cite{Wang2021}  \\
 & $\mathrm{NbSe_2}$&Ohmic&2.02 &-0.01&4.68&1.88&1.55&3.41&-&-&\cite{Liang} \\
 & $\mathrm{TaS_2}$&Ohmic&2.18&-0.05&5.02&1.90&1.28&3.89&-&-&\cite{Liang} \\
 & $\mathrm{TaSe_2}$&Schottky&1.98&0.02&4.60&1.87&1.65&3.28&-&-&\cite{Liang} \\
 & Graphene &Schottky&1.56&0.78&4.04&1.89&2.05&3.00&-&-&\cite{Liang}\\
 &&Schottky&1.79 &0.22&4.39&1.89&1.73&3.26&-&-&\cite{Wang2021} \\
 &&Schottky&1.67&0.85&0&0&100&0&\checkmark&\ding{55}&\cite{Ma2023} \\
 \hline
 
$\mathrm{MoGeSiN_{4}}$ & Graphene &Schottky&0.63&1.13&-&-&-&-&\checkmark&\checkmark&\cite{Binh2021}\\
$\mathrm{MoSiGeN_{4}}$ & Graphene &Schottky&1.03&0.74&-&-&-&-&\checkmark&\checkmark&\cite{Binh2021} \\
\hline

$\mathrm{MoSi_2P_4}$&Graphene&Schottky&0.13&0.60&6.11&1.04&7.08&0.77&\checkmark&\checkmark& \cite{Li2023_graphene}\\
\hline
$\mathrm{MoSi_2As_4}$&Graphene&Schottky&0.32&0.31&5.67&1.31&4.02&1.14&\checkmark&\checkmark&\cite{Li2023_graphene} \\
\hline

$\mathrm{CrSi_2N_4}$&Graphene&Schottky&0.03&0.45&4.35&1.89&1.77&3.226&-&-&\cite{Shu2023} \\
\hline
$\mathrm{CrC_2N_4}$&Graphene&Schottky&0.05&1.85&4.08&1.69&3.02&2.046&-&-&\cite{Shu2023} \\
\hline
\hline

\end{tabular}
}
\end{table*}

\begin{table*}
\caption{\label{MS_3D_table}Key interfacial properties of 3D/2D MS contacts composed of MA$_2$Z$_4$.} 

\centering
 \resizebox{\textwidth}{!}
{
\begin{tabular}{>{\centering\arraybackslash}m{1.5cm}>{\centering\arraybackslash}m{1cm}>{\centering\arraybackslash}m{2cm}>{\centering\arraybackslash}m{2cm}>{\centering\arraybackslash}m{2cm}>{\centering\arraybackslash}m{1.5cm}>{\centering\arraybackslash}m{2cm}>{\centering\arraybackslash}m{2cm}>{\centering\arraybackslash}m{2cm}>
{\centering\arraybackslash}m{1.5cm}}
\hline
\hline

 $\mathrm{\bf{MA_{2}Z_{4}}}$ & \bf{Metal} & \bf{Contact Type} &  \multicolumn{2}{c}{\bf{Schottky Barrier Height (eV)}} &\bf{Tunnelling Barrier Height (eV)} & \bf{Tunnelling Barrier Width (Å)} & \bf{Tunnelling Barrier Probability} & \bf{Tunnelling Specific Resistivity (10$^{-9}$ 
  $\Omega$ cm$^{2}$)} & \bf{Ref.} \\
 
&&&n-type&p-type&&&&& \\
 \hline

\multirow{9}{*}{$\mathrm{MoSi_{2}N_{4}}$}
& Sc &Ohmic&-0.06&2.03&0&0&100&0&\cite{Wang2021} \\
 & Ti &Ohmic&-0.12&2.06 &0&0&100&0&\cite{Wang2021} \\
 & Ni &Schottky&0.82&1.13 &2.08&1.29&14.86&1.27$\times10^{-1}$&\cite{Wang2021}\\
 & Cu &Schottky&0.65&1.48 &2.80&1.23&12.14&3.71&\cite{Wang2021} \\
 & Pd &Schottky&1.14&0.97 & 2.98 &1.13&13.55&1.02$\times10^{-1}$&\cite{Wang2021} \\
 & Ag &Schottky&0.67&1.37 &3.01&1.42&8.01&1.56&\cite{Wang2021} \\
 & In &Schottky&0.43&1.59 &2.94&1.69&5.13&1.86&\cite{Wang2021} \\
 & Pt &Schottky&1.85&0.20 &4.04&1.46&4.94&1.39&\cite{Wang2021} \\
 & Au &Schottky&1.44&0.68 &3,73&1.48&5.35&1.43&\cite{Wang2021} \\
 \hline
 
\multirow{9}{*}{$\mathrm{WSi_{2}N_{4}}$}
& Sc &Ohmic&-0.06&2.03&0&0&100&0&\cite{Wang2021} \\
 & Ti &Ohmic&-0.12&2.06 &0&0&100&0&\cite{Wang2021} \\
 & Ni &Schottky&0.82&1.13 &2.08&1.29&14.86&1.27$\times10^{-1}$&\cite{Wang2021}\\
 & Cu &Schottky&0.65&1.48 &2.80&1.23&12.14&3.71&\cite{Wang2021} \\
 & Pd &Schottky&1.14&0.97 & 2.98 &1.13&13.55&1.02$\times10^{-1}$&\cite{Wang2021} \\
 & Ag &Schottky&0.67&1.37 &3.01&1.42&8.01&1.56&\cite{Wang2021} \\
 & In &Schottky&0.43&1.59 &2.94&1.69&5.13&1.86&\cite{Wang2021} \\
 & Pt &Schottky&1.85&0.20 &4.04&1.46&4.94&1.39&\cite{Wang2021} \\
 & Au &Schottky&1.44&0.68 &3,73&1.48&5.35&1.43&\cite{Wang2021} \\ 
 \hline

 \multirow{7}{*}{$\mathrm{CrSi_{2}N_{4}}$}
&Ag&Ohmic&-0.03&0.54&2.91&1.3&10.45&1.954&\cite{Shu2023} \\
&Au&Schottky&0.02&0.54&3.69&1.44&5.86&1.372&\cite{Shu2023} \\
&Cu&Ohmic&-0.03&0.52&2.79&1.09&15.48&8.9$\times10^{-2}$&\cite{Shu2023} \\
&Ni&Ohmic&-0.01&0.6&3.17&1.2&11.27&2.205&\cite{Shu2023} \\
&Pd&Ohmic&-0.02&0.74&2&0.8&31.42&3.81$\times10^{-2}$&\cite{Shu2023} \\
&Pt&Ohmic&-0.01&0.75&2.99&1.08&14.76&8.93$\times10^{-2}$&\cite{Shu2023} \\
&Ti&Ohmic&-0.04&0.58&2.04&1.03&22.21&6.87$\times10^{-2}$&\cite{Shu2023} \\
\hline

\multirow{7}{*}{$\mathrm{CrC_{2}N_{4}}$}
&Ag&Schottky&0.07&1.84&3.52&1.51&5.55&1.499&\cite{Shu2023} \\
&Au&Schottky&0.32&1.54&4.1&1.55&4.01&1.601&\cite{Shu2023} \\
&Cu&Ohmic&-0.06&1.79&1.24&0.6&50.48&2.3$\times10^{-2}$&\cite{Shu2023} \\
&Ni&Schottky&0.15&1.81&0.32&0.13&88.74&5.23$\times10^{-3}$&\cite{Shu2023} \\
&Pd&Schottky&0.14&1.81&3.64&1.3&7.84&1.294&\cite{Shu2023} \\
&Pt&Schottky&0.19&1.73&3.69&1.22&9.07&1.340&\cite{Shu2023} \\
&Ti&Ohmic&-0.13&1.75&0&0&100&0&\cite{Shu2023} \\

\hline
\hline
\end{tabular}
}
\end{table*}

\begin{table*}
\caption{\label{SS_table} Key quantities and the representative applications of SS vdWHs. When available, both PBE and HSE calculated bandgaps are included. The subscripts $\uparrow$ and $\downarrow$ denote the band gap of spin-up and spin-down subband, respectively. For quantities not reported, the symbol `-' is used.}
\centering
 \resizebox{\textwidth}{!}
 {
\begin{tabular}{>{\centering\arraybackslash}m{1.5cm}>{\centering\arraybackslash}m{2cm}>{\centering\arraybackslash}m{2cm}>{\centering\arraybackslash}m{3.2cm}>{\centering\arraybackslash}m{2.2cm}>{\centering\arraybackslash}m{1cm}>{\centering\arraybackslash}m{1cm}>{\centering\arraybackslash}m{1cm}>{\centering\arraybackslash}m{2cm}>{\centering\arraybackslash}m{3cm}>{\centering\arraybackslash}m{1cm}}

\hline
\hline

 $\mathrm{\bf{MA_{2}Z_{4}}}$ & \bf{Semiconductor} & \bf{Alignment Type} &  \multicolumn{2}{c}{\bf{Bandgap (eV)}} &\multicolumn{3}{c}{\bf{Bandgap Tunability}}& \bf{Optical Absorption Peaks} & \bf{Application} & \bf{Ref.} 
 
 \\ 
 

&&& Magnitude & Direct / Indirect & Electric Field & Vertical Strain & Biaxial Strain&&& \\
 \hline
 
\multirow{14}{*}{$\mathrm{MoSi_{2}N_{4}}$}
  & $\mathrm{MoS_2}$&Type-II&1.26 (PBE), 1.84 (HSE)&Indirect&-&-&-&UV&Optoelectronics&\cite{Bafekry2021} \\
  &&Type-II&1.12 (PBE) &Indirect&\checkmark&-&\checkmark&UV&Optoelectronics&\cite{Xu2023} \\
  &&Type-II&1.93 (HSE)&Indirect&-&-&-&UV&H$_2$O Photocatalyst& \cite{Jalil2023} \\
  & $\mathrm{MoSe_2}$&Type-I& 1.39 (PBE), 1.82 (HSE) &Direct&\checkmark&-&\checkmark&Visible, UV &FET, Optoelectronics&\cite{Cai2021_MoSe2} \\
& $\mathrm{WSe_2}$&Type-I &1.17 (PBE), 1.57 (HSE)&Direct&\checkmark&-&-&Visible, UV&FET, Photovoltaic&\cite{Cai2021_WSe2}\\
&&Type-II&1.38(PBE),1.81(HSE)&Indirect&-&-&-&Visible, UV&H$_2$O Photocatalyst&\cite{Liu2023_WSe2}\\
 & InSe&Type-II&1.35 (PBE), 1.61 (HSE)&Direct&-&-&-&UV&$\mathrm{H_2O}$ Photocatalyst&\cite{He2022}\\
 & ZnO&Type-II &1.60 (PBE)& Indirect&\checkmark&\checkmark&-&-&Optoelectronics&\cite{JinQuan2022}\\
 & GaN &Type-I&1.56 (PBE)&Direct &\checkmark&\checkmark&-&-&Optoelectronics&\cite{JinQuan2022}\\
 & $\mathrm{C_{2}N}$ &Type-II&1.74 (HSE)&Direct&-&-&-&Visible, UV&$\mathrm{H_2O}$ Photocatalyst&\cite{Zeng2021}\\
 & $\mathrm{C_3N_4}$ &Type-II&1.86 (PBE), 2.66 (HSE)&Direct&\checkmark&\checkmark&\checkmark&Visible, UV&Optoelectronics&\cite{Nguyen2022_C3N4}\\
 & $\mathrm{CrCl_3}$ &Type-II&$0.5_{\uparrow}$, $1.67_{\downarrow}$&Indirect&-&\checkmark&-&-&Valleytronics&\cite{Zhao2021}\\
 & $\mathrm{Cs_3Bi_2I_9}$ &Type-II&0.91 (PBE)&Direct&-&\checkmark&\checkmark&Visible, UV&Optoelectronics&\\
 &&&&&&&&& $\mathrm{H_2O}$ Photocatalyst&\cite{Liu2022}\\
 
 & BlueP &Type-II&1.13 (PBE), 2.02 (HSE)&Indirect &\checkmark&-&\checkmark&Visible, UV&
 FET, Optoelectronics&\cite{Fang2022}\\
 &&Type-II&1.18(PBE), 2.00(HSE)&Indirect&-&-&\checkmark&Visible, UV& $\mathrm{H_2O}$ Photocatalyst&\cite{Xuefeng2022}\\
 
 &$\mathrm{CrS_2}$&Type-II&0.93 (PBE), 1.52 (HSE)&Indirect&-&\checkmark&\checkmark&Visible, UV&$\mathrm{H_2O}$ Photocatalyst&\cite{Li2022}\\

&SeMoS&Type-I&1.55 (HSE)&Indirect&-&-&-&Visible, UV&Optoelectronics&\cite{Jalil2023} \\
 
 & $\mathrm{WSi_2N_4}$&Type-II & 1.70 (PBE), 2.15(HSE) &Indirect&-&-&-&Visible, UV &$\mathrm{H_2O}$ Photocatalyst&\cite{Zhao2022}\\
 \hline

\multirow{2}{*}{$\mathrm{\alpha_2-MoSi_2N_4}$}&MoS$_2$&Type-II&2.08 (HSE)&Indirect&-&-&-&UV&H$_2$O Photocatalyst&\cite{Jalil2023} \\
&SeMoS&Type-II&1.26 (HSE)&Indirect&-&-&-&Visible, UV&H$_2$O Photocatalyst&\cite{Jalil2023} \\
\hline

 \multirow{2}{*}{$\mathrm{MoGe_{2}N_{4}}$}
 & SMoTe&Type-I&0.27 (PBE)&Direct&-&\checkmark&\checkmark&Visible, UV&Optoelectronics&\cite{Wang2021_MoGe2N4} \\
  & TeMoS&Type-III&0 (PBE)&NA&-&\checkmark&\checkmark&Visible, UV&Optoelectronics&\cite{Wang2021_MoGe2N4} \\
 \hline
 
 $\mathrm{MoSi_2P_4}$ & BP & Type-II &1.02 (HSE)&Direct&-&-&-&Visible, UV&Photovoltaic&\cite{Guo2022_BP}\\ 
 \hline

\multirow{2}{*}{$\mathrm{MoSiGeN_{4}}$}&\multirow{2}{*}{SeMoS}&\multirow{2}{*}{Type-II}& \multirow{2}{*}{0.87 (PBE),1.28 (HSE)}&\multirow{2}{*}{Direct}&-&\multirow{2}{*}{\checkmark}&\multirow{2}{*}{\checkmark}&Visible&\multirow{2}{*}{Photovoltaic} & \multirow{2}{*}{\cite{Zhang2023}} \\
&&&&&&&&Infrared&& \\
\hline

\multirow{2}{*}{$\mathrm{WSi_{2}N_{4}}$}&MoSe$_2$&Type-II&1.26 (PBE)&Direct&\checkmark&\checkmark&\checkmark&Visible, UV&Optoelectronics&\cite{Pei2023} \\
&WSe$_2$&Type-II&1.16 (PBE)&Direct&\checkmark&\ding{55}&\checkmark&Visible, UV&Optoelectronics&\cite{Pei2023} \\

 \hline 
 \hline
 
\end{tabular}
}
\end{table*}


\subsection{\label{Photocatalyst & Photovoltaic}Photocatalyst \& Solar Energy Conversion}

\subsubsection{Photocatalytic water splitting}
Photocatalytic water splitting (PCWS) is an effective way to solve the problem of energy crisis and environmental pollution. PCWS usually consists of the following steps \cite{Wang2022_catalyst, Fu2018}: (1) Photons with energy larger than the photocatalyst's bandgap excite electrons from its valence bands to conduction bands, generating electron-hole pairs; (2) photogenerated carriers diffuse or drift to the surfaces of the photocatalyst; and (3) photogenerated electrons participate in the hydrogen evolution reaction (HER) to generate hydrogen, whereas photogenerated holes participate in the oxygen evolution reaction (OER) to generate oxygen. The conventional PCWS photocatalyst should meet the following requirements to enable the two half-reactions to happen simultaneously \cite{Chen2017,Kudo2009}: (i) photocatalyst having an appropriate bandgap to obtain sufficient light harvesting; (ii) band edge positions of photocatalyst should span the $\mathrm{H_2O}$ redox potential levels; and (iii) the photogenerated carriers should provide sufficient driving force to overcome the energy barriers in HER and OER processes, i.e. the potentials between the photocatalyst's CBM and the $\mathrm{H_2O}$ reduction level must be large enough to provide strong reduction ability, while the potential between photocatalyst's VBM and the $\mathrm{H_2O}$ oxidation level should be large enough to provide strong oxidation ability. 2D photocatalysts have drawn extensive attentions as they exhibit obvious advantages in improving the photocatalytic activities \cite{Liu2018, Su2018}, such as tuning their bandgaps through altering the thickness, introducing atomic defects, or applying strain. The ultrathin thickness of 2D materials could also shorten the photogenerated carriers' migration distances, and their planar structures with high specific surface areas can provide more active sites and offer ideal platforms for combining with other co-catalysts. 

\begin{figure}[t]
\includegraphics[width=0.5\textwidth]{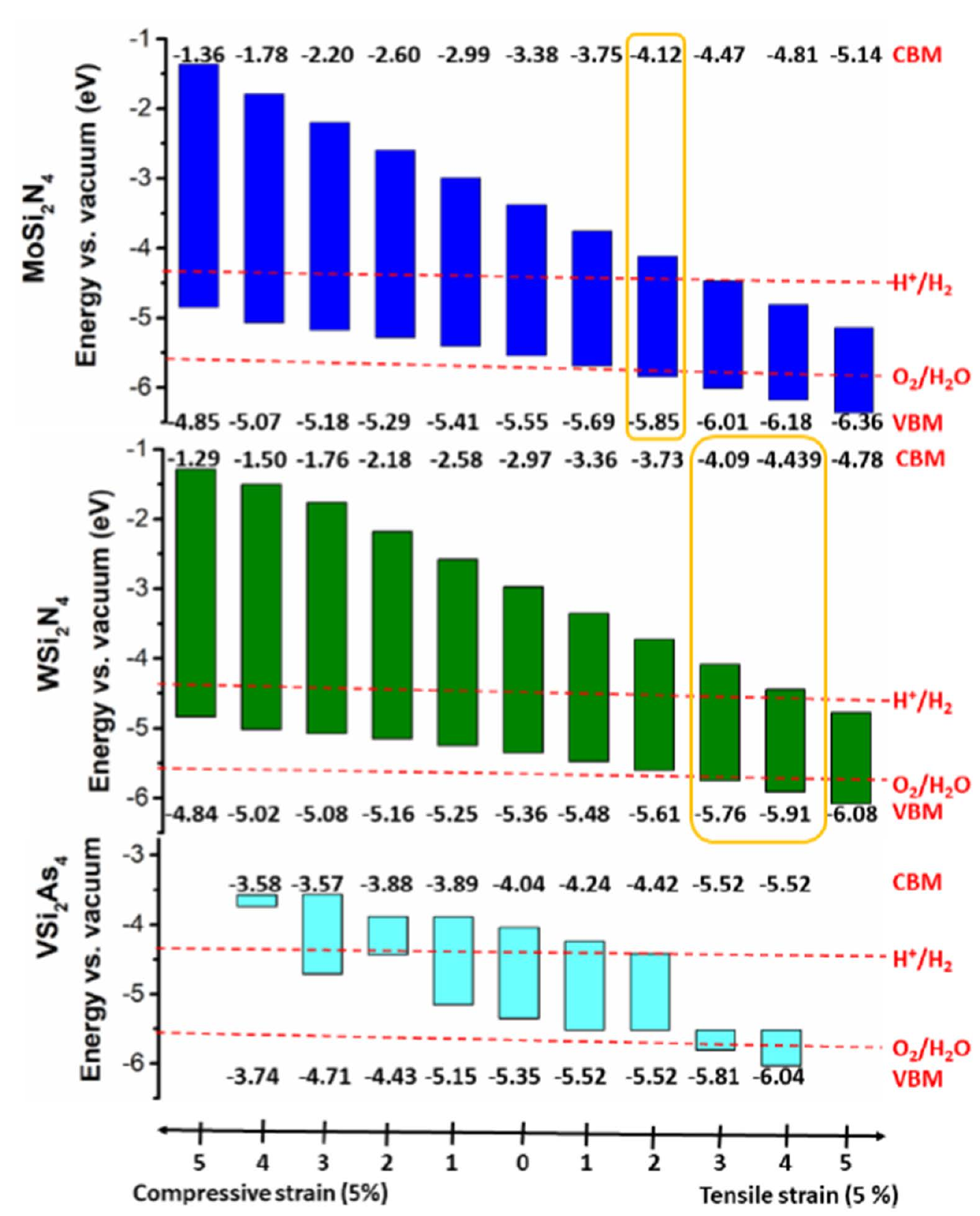}
\caption{\label{Fig8} \textbf{Photocatalytic water splitting applications of MA$_2$Z$_4$.} Band alignments of $\mathrm{MoSi_2N_4}$, $\mathrm{WSi_2N_4}$ and $\mathrm{VSi_2As_4}$ with respect to redox potential of water under different applied biaxial strain. Yellow boxes signify the band edges straddles the $\mathrm{H_2O}$ redox potentials. Reproduced with permission from Chen \& Tang, Chem. Eur. J. \textbf{27}, 9925 (2021), Copyright 2021 Wiley‐VCH GmbH.}
\end{figure} 

\begin{figure*}[t]
\includegraphics[width=1\textwidth]{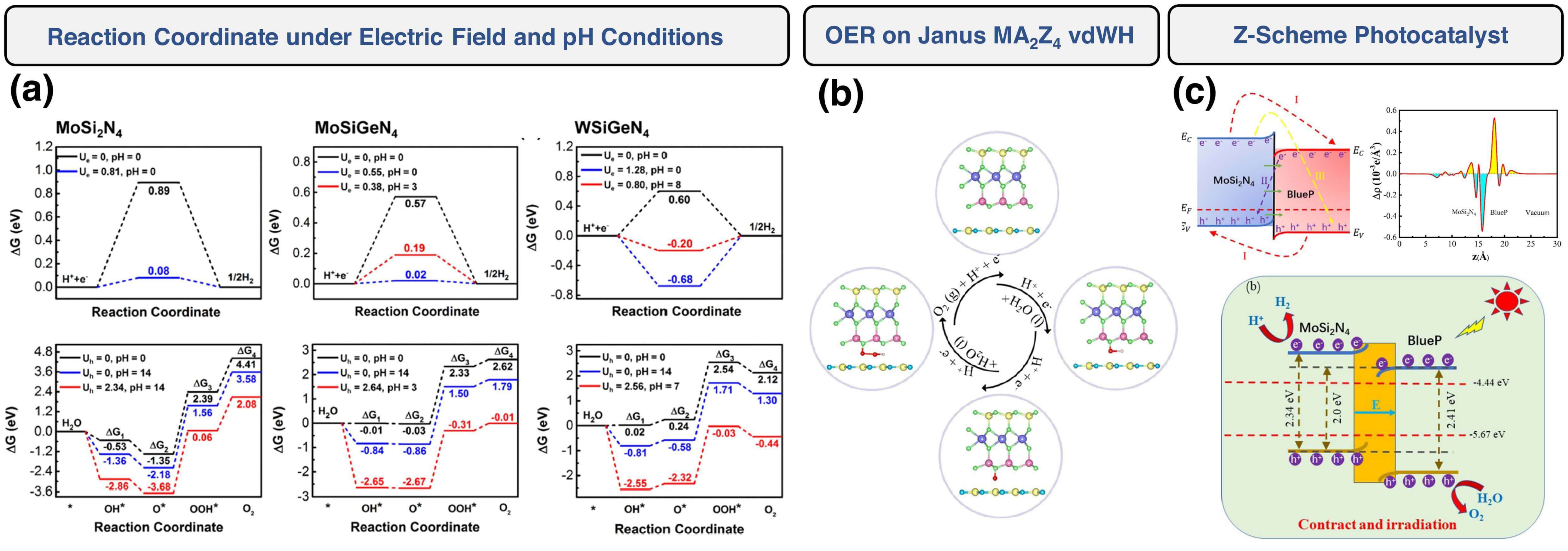}
\caption{\label{Fig9} \textbf{Photocatalytic water splitting applications of MA$_2$Z$_4$-based Heterostructures.} (a) Free energy diagrams at each reaction stage of HER and OER of (Left) $\mathrm{MoSi_2N_4}$, (Middle) $\mathrm{WSi_2N_4}$ and (Right) Janus-$\mathrm{WSiGeN_4}$ at different pH levels with/without light illumination. 
(b) The four-step reaction of OER in Janus-$\mathrm{MoSiGeN_4/SiC}$ vdWH. 
(c) $\mathrm{MoSi_2N_4/BlueP}$ is found to be a Z-scheme water-splitting photocatalyst: (Top Left) Schematic of band bending at the interface and paths that photoexcited charge carriers can follow, (Top Right) plane-average charge density profile along the stacking axis showing regions of electron accumulation (yellow) and depletion (light blue) resulting in an intrinsic electric field, (Bottom) Schematic of the redox reactions taking place in the Z-scheme photocatalyst. (a) Reproduced with permission from Yu $et \ al.$, ACS Appl. Mater. Interfaces \textbf{13}, 2809 (2021), Copyright 2021 American Chemical Society. (b) Reproduced with permission from Huang $et \ al.$, J. Mater. Sci. \textbf{57}, 16404 (2022), Copyright 2022 Springer Nature. (c) Reproduced with permission from Xuefeng $et \ al.$, J. Phys. D: Appl. Phys. \textbf{55}, 215502 (2022), Copyright 2022 IOP Publishing Ltd.}
\end{figure*}

The tunability of the bandgap of $\mathrm{MA_2Z_4}$ family for use as photocatalysts motivates further exploration of their potential PCWS ability. By exploring the electronic and photocatalytic features of $\mathrm{MA_2Z_4}$ (M = Cr, Mo, W; A = Si, Ge; Z = N, P) monolayers, Mortazavi $et \ al.$ \cite{Mortazavi2021} found that the $\mathrm{MoSi_2N_4}$ and $\mathrm{WSi_2N_4}$ monolayers possess suitable band edge positions, high electron and hole mobility, and strong visible light harvesting ability, which makes these monolayers highly promising candidates for PCWS applications. By screening 54 kinds of possible combinations of $\mathrm{MA_2Z_4}$ family, Chen $et \ al.$ have confirmed that only $\mathrm{MoSi_2N_4}$ and $\mathrm{WSi_2N_4}$ under feasible biaxial tensile strain possess the ability for overall PCWS [see Fig. \ref{Fig8}] \cite{Chen2021}. Theoretical study by Jian $et \ al.$ \cite{Jian2021} reported that pristine and compressive $\mathrm{MoSi_2N_4}$ are promising PCWS photocatalysts having excellent solar energy absorption abilities for water oxidation and reduction. Particularly, compressive strain leads to the spatial separation of photoexcited electrons and holes, which is favorable for enhancing their separation rates. Rehman $et \ al.$ explored the electronic and photocatalytic properties of $\mathrm{MC_2N_4}$ (M = Cr, Mo, W) monolayers based on bandgap center \cite{Rehman2022}. These three monolayers possess sufficient kinetic potentials to drive both water redox reactions, with in-plane compressive strains found to enhance the  overall PCWS of all these monolayers. However, pristine and biaxially strained $\mathrm{CrC_2N_4}$ is more favourable for PCWS application due to its strong absorption in the visible regime as compared to $\mathrm{MoC_2N_4}$ and $\mathrm{WC_2N_4}$ which have strong absorption ability only in the ultraviolet regime. Ren $et \ al.$ evaluated the HER performances of MX$_2$Y$_4$ monolayers (M = Cr, Hf, Mo, Ti, W, Zr; X = Si, Ge; Y = N, P, As) and their mechanical properties, unearthing TiSi$_2$N$_4$ and ZrSi$_2$N$_4$ as highly efficient HER catalysts due to the small change in Gibbs free energy during $\mathrm{H^{+}}$ adsorption \cite{Ren2023}.

By using high-throughput screening, Lin $et \ al.$ unearthed 35 $\mathrm{MA_2Z_4}$-based monolayers with appropriate bandgap and suitable band edge positions from 648 candidates \cite{Lin2022}. After further analysis of the structural stability, free energy change in OER, exciton property, and carrier mobility, $\mathrm{\beta_2}$-$\mathrm{HfSi_2N_4}$ and $\mathrm{\beta_2}$-$\mathrm{ZrSi_2N_4}$ have been chosen as highly efficient OER photocatalysts. The bandgaps of $\mathrm{MA_2Z_4}$ monolayers are found to primarily dependent on their M-Z bonding strength and electron distribution, while their OER activities are closely related to the adsorption ability of the O atom on the reaction site. 
It is worthwhile to note that the potential of MA$_2$Z$_4$ family for CO$_2$ reduction has also been investigated, such as the lateral $\mathrm{(MoSi_2N_4)_{5-n}/(MoSiGeN_4)_n}$. High-throughput screening on 104 types of Janus-MA$_2$Z$_4$ monolayers has uncovered $\mathrm{MoSiGeP_4}$ and $\mathrm{HfSiGeP_4}$ as efficient CO$_2$ photocatalysts in the visible regime \cite{Zhang2022_janus}.

\textbf{Role of electric field in PCWS.} Recently, PCWS reactions based on Janus 2D materials have been proposed \cite{Yang2014,Fu2018_intrinsic_efield}. The intrinsic electric field in Janus 2D materials not only promotes the spatial separation of electrons and holes but also breaks the conventional bandgap (1.23 eV) restriction required of PCWS. This enables the harvesting of photons with frequencies below the visible regime for PCWS. Inspired by the first synthesis of Janus-MoSSe, the photocatalytic performance of other Janus materials have been explored \cite{Peng2019,Ma2018,Ju2020,Luo2021}. Yu $et \ al.$ predicted the excellent stability of Janus-$\mathrm{MoSiGeN_4}$ and Janus-$\mathrm{WSiGeN_4}$ monolayers for PCWS reactions, in which both of these two monolayers are semiconductors that have indirect bandgaps, suitable band edge positions, excellent optical absorption ability, and large carrier mobility anisotropy \cite{Yu2021}. The intrinsic electric fields induced by these Janus structures not only boost the spatial separation of photogenerated carriers, but also provides a potential offset at the HER and OER catalytic sites which enables the photogenerated carriers to provide additional driving force to lower the Gibbs free energy in the redox reactions [see Fig. \ref{Fig9}(a)]. Additionally, introducing N atom vacancies to the surface effectively lowers the HER and OER overpotentials. When pH is 3 or $\leq 8$, the overall $\mathrm{H_2O}$ redox processes could proceed simultaneously and spontaneously on the surfaces of Janus-$\mathrm{MoSiGeN_4}$ and Janus-$\mathrm{WSiGeN_4}$ under light irradiation [see Fig. \ref{Fig9}(b) for a schematic of OER reaction taking place on Janus-$\mathrm{MoSiGeN_4}$]. Lv $et \ al.$ investigated the structure, electronic property, and carrier mobility of various stacking configurations of bilayer Janus-$\mathrm{MoSiGeN_4}$ \cite{Lv2022}. The dipole moment formed across the structure plays a big part in modulating the bandgaps through creating a built-in electric field, which causes a shift in the energy bands of each monolayer to achieve Type-II alignment. Additionally, the carrier mobility of bilayer Janus-$\mathrm{MoSiGeN_4}$ ($\mathrm{58522.3 \ cm^{2}V^{-1}s^{-1}}$) is enahcned by one order of magnitude compared to its monolayer form ($\mathrm{4883.23 \ cm^{2}V^{-1}s^{-1}}$), thus revealing Janus-$\mathrm{MoSiGeN_4}$ as a multifunctional 2D materials that also has potential use in electronics device applications.

\textbf{PCWS based on MA$_2$Z$_4$ vdWHs.} The construction of vdWHs offer another effective avenue to design highly efficient PCWS catalysts \cite{Moniz2015,Wang2014}. Based on the charge transfer path, the mechanism of PCWS catalysts in vdWHs can usually be divided into either Type-II or Z-scheme \cite{Zhou2021}. Since the first successful synthesis of $\mathrm{MoSi_2N_4}$, some $\mathrm{MoSi_2N_4}$-based Type-II and Z-scheme vdWH photocatalysts are proposed. Zeng $et \ al.$ reported the good thermodynamic stability and potential PCWS application of $\mathrm{MoSi_2N_4/C_2N}$ \cite{Zeng2021} which has efficient charge carrier separation at the interface and excellent optical absorption in the visible regime. Moreover the absolute value of its free energy in hydrogen adsorption is very close to zero. The $\mathrm{MoSi_2N_4/C_2N}$ thus exhibits excellent HER ability for $\mathrm{H_2}$ gas generation. In addition, other $\mathrm{MoSi_2N_4}$-based vdWHs such as $\mathrm{MoSi_2N_4/WSe_2}$ \cite{Liu2023_WSe2}, $\mathrm{MoSi_2N_4/MoSe_2}$ \cite{Cai2021_MoSe2}, $\mathrm{MoSi_2N_4/CrS_2}$ \cite{Li2022}, $\mathrm{MoSi_2N_4/InSe}$ \cite{He2022}, $\mathrm{MoSi_2N_4/WSi_2N_4}$ \cite{Zhao2022} are predicted to be potential vertical PCWS catalysts.  

$\mathrm{MoSi_2N_4/Blue Phosphorus}$ (BlueP) vdWH has potential in PCWS due to its Type-II band alignment \cite{Xuefeng2022}. There are three possible photoinduced carrier transfer paths in $\mathrm{MoSi_2N_4/BlueP}$ [see Fig. \ref{Fig9}(c)(Top Left)]: (Path-I) electrons transfer from the $\mathrm{MoSi_2N_4}$’s conduction bands to the BlueP’s valence bands and the $\mathrm{MoSi_2N_4}$’s valence bands, (Path-II) electrons and holes recombine with each other between the BlueP’s conduction bands and the $\mathrm{MoSi_2N_4}$’s valence bands, and (Path-III) electrons transfer from $\mathrm{MoSi_2N_4}$’s conduction bands to the BlueP’s valence bands. The presence of a built-in electric field created by the interface dipole [see Fig. \ref{Fig9}(c)(Top Right)] leads to the suppression of charge recombination through Path-I and Path-III. When the charges recombine along Path-II, the excess electrons at the $\mathrm{MoSi_2N_4}$’s conduction bands participate in the HER and the excess holes at the BlueP’s valence bands participates in the OER [see Fig. \ref{Fig9}(c)(Bottom)]. The energy difference between $\mathrm{MoSi_2N_4}$’s CBM and $\mathrm{H^{+}/H_{2}}$ HER potential, and the energy difference between BlueP’s VBM and $\mathrm{H_2O/O_2}$ OER potential at pH = 0, are 0.71 and 0.81 eV, respectively. These values are higher than the assumed energy loss in carrier migration of 0.2 in HER and 0.6 eV in OER \cite{Fu2018_intrinsic_efield}, implying that $\mathrm{MoSi_2N_4/BlueP}$ vdWH has both good reduction and oxidation ability. Moreover, its light absorption performance could be further improved by applying tensile strain in the range of 0 to 8\%. These findings show that $\mathrm{MoSi_2N_4/BlueP}$ vdWH is a promising Z-scheme PCWS catalyst. Huang $et \ al.$ theoretically screened 4 configurations of Janus-$\mathrm{MoSiGeN_4/SiC}$ vdWHs as highly efficient Z-scheme PCWS catalysts from a total of 12 stacking configurations \cite{Huang2022}. 
This suggests that stacking configurations and twist engineering \cite{Cao2018_twist} can potentially be used to further optimize the PCWS capabilities of MA$_2$Z$_4$ heterostructures. $\mathrm{MoSi_2N_4/MoSX}$ (X = S, Se) has been studied for their HER ability with the introduction of vacancy defects, showing that these vdWHs are Z-scheme PCWS photocatalysts, and S/Se vacancies are able to provide the most favourable vacant sites for the H atom to undergo HER at pH = 0 \cite{Jalil2023}.  

\begin{figure}[t]
\includegraphics[width=0.485\textwidth]{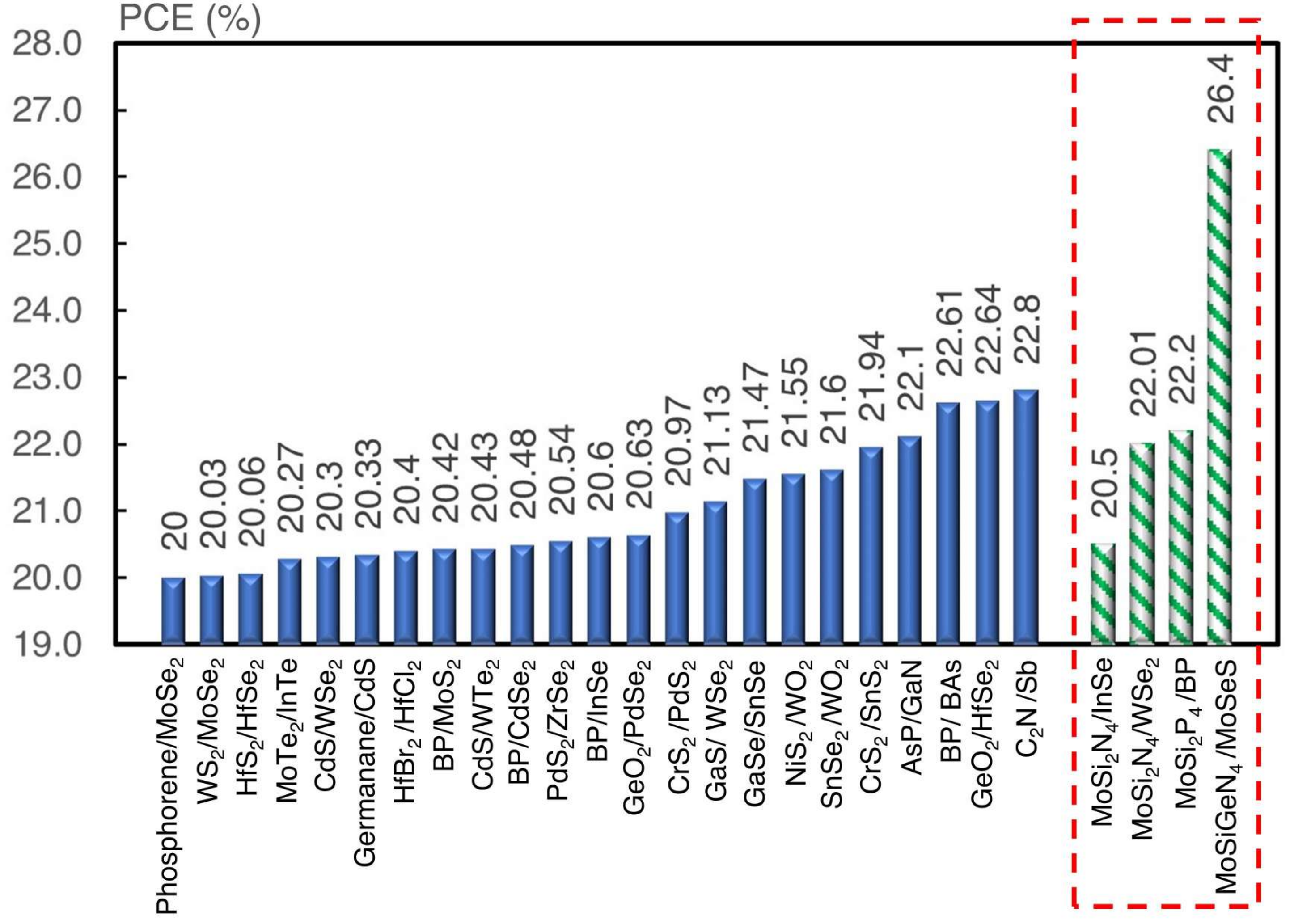}
\caption{\label{Fig10} \textbf{Excitonic solar cell maximum power conversion efficiency (PCE) of various Type-II vdWHs}. vdWHs with PCE greater than 20\% are listed for comparison. Data taken from Ref. \cite{Tho2022, Guo2022_BP, Linghu2018, Kaur2018, Huang2020,Mohanta2019,Mohanta2020,Mao2019,Xie2016,Wang2018_C2N, Zhang2023}.}
\end{figure}

\subsubsection{Solar-to-\textcolor{black}{electricity} energy conversion} 
$\mathrm{MA_2Z_4}$-based heterostructures have attracted much attention in solar cell applications due to the generally good spatial separation of photogenerated carriers in vdWHs \cite{Liu2021_charge}. For example, $\mathrm{MoSi_2P_4/BP}$ vdWH \cite{Guo2022_BP} has can effectively separate the photogenerated carriers due to its Type-II. The hole carrier mobility and optical absorption ability of $\mathrm{MoSi_2P_4/BP}$ are improved as compared to its constituent monolayers. Moreover, the predicted solar-to-electricity power conversion efficiency (PCE) based on an excitonic solar cell setup reaches 22.2\%, which is comparable to other high-performance (PCE $>$ 15\%) vdWHs obtained through high-throughput screening \cite{Linghu2018} and in several other works \cite{Mohanta2019, Mohanta2020, Mao2019, Xie2016, Wang2018_C2N, Huang2020, Tho2022}. Janus-MoSSe/MoGeSiN$_4$ is found to have multiple strain-tunable absorption peaks in the visible and infrared regime, and exhibits high PCE value of $>$ 27\% through changing its interlayer distance and biaxial strain \cite{Zhang2023}.
$\mathrm{MoSi_2N_4/InSe}$ and $\mathrm{MoSi_2N_4/WSe_2}$ are also potential candidates for excitonic solar cell with PCE of 20.5\% and 22.0\%, respectively. Fig. \ref{Fig10} shows the reported PCE values of MA$_2$Z$_4$-based vdWHs benchmarked against other high solar-to-electricity performance vdWHs that are potential candidates for constructing excitonic solar cells. 
\textcolor{black}{
We note that solar cell efficiency well over 20\% has been achieved in other materials, such as crystalline silicon and perovskite \cite{correa2017promises, mcmeekin2019solution,tu2021perovskite}. 
The MA$_2$Z$_4$ heterostructure-based solar cell has several potential advantageous over bulk semiconductor and perovskites. For instance, vdWH engineering of MA$_2$Z$_4$ and other 2D materials can be used to broaden the absorption spectral range while still retaining their ultracompact device geometry, an aspect difficult to achieve using bulk semiconductors. The excellent ambient stability of MA$_2$Z$_4$ is also more advantageous than other nanomaterial-based solar cell, such as perovskites and organic solar cells \cite{chen2019organic}, which are faced with long-term stability issues. 
}

In $\mathrm{MoSi_2N_4/C_3N_4}$ heterostructure, a direct bandgap semiconductor with Type-II band alignment, possesses strong visible light absorption \cite{Nguyen2022_C3N4}. The band alignments could be switched between Type-I and Type-II by using external electric field, modulation of interlayer spacing, and in-plane strain [see Fig. \ref{Fig5}(b)-(d)] which renders $\mathrm{MoSi_2N_4/C_3N_4}$ a promising candidate as a tunable optoelectronic materials. The contact type of $\mathrm{MoSi_2N_4/MoS_2}$ can also be tuned by the same methods and a photodetector based on the vdWH has been proposed, which exhibits an enahcned the photocurrent when a tensile strain is applied across the device \cite{Xu2023}, suggesting potential coupled opto-electro-mechanical applications \cite{midolo2018nano}. Other $\mathrm{MA_2Z_4}$-based vdWHs [see Fig. \ref{Fig5}(b)-(d)] have also been shown to possess tunable band alignment such as $\mathrm{MoSi_2N_4/Cs_3Bi_2I_9}$ \cite{Liu2022}, $\mathrm{MoSi_2N_4/WSe_2}$ \cite{Cai2021_WSe2}, $\mathrm{MoSi_2N_4/GaN}$ \cite{JinQuan2022}, $\mathrm{MoSi_2N_4/ZnO}$ \cite{JinQuan2022} and Janus-$\mathrm{MoSTe/MoGe_2N_4}$ \cite{Wang2021_MoGe2N4}. 

\begin{figure}[t]
\includegraphics[width=0.4\textwidth]{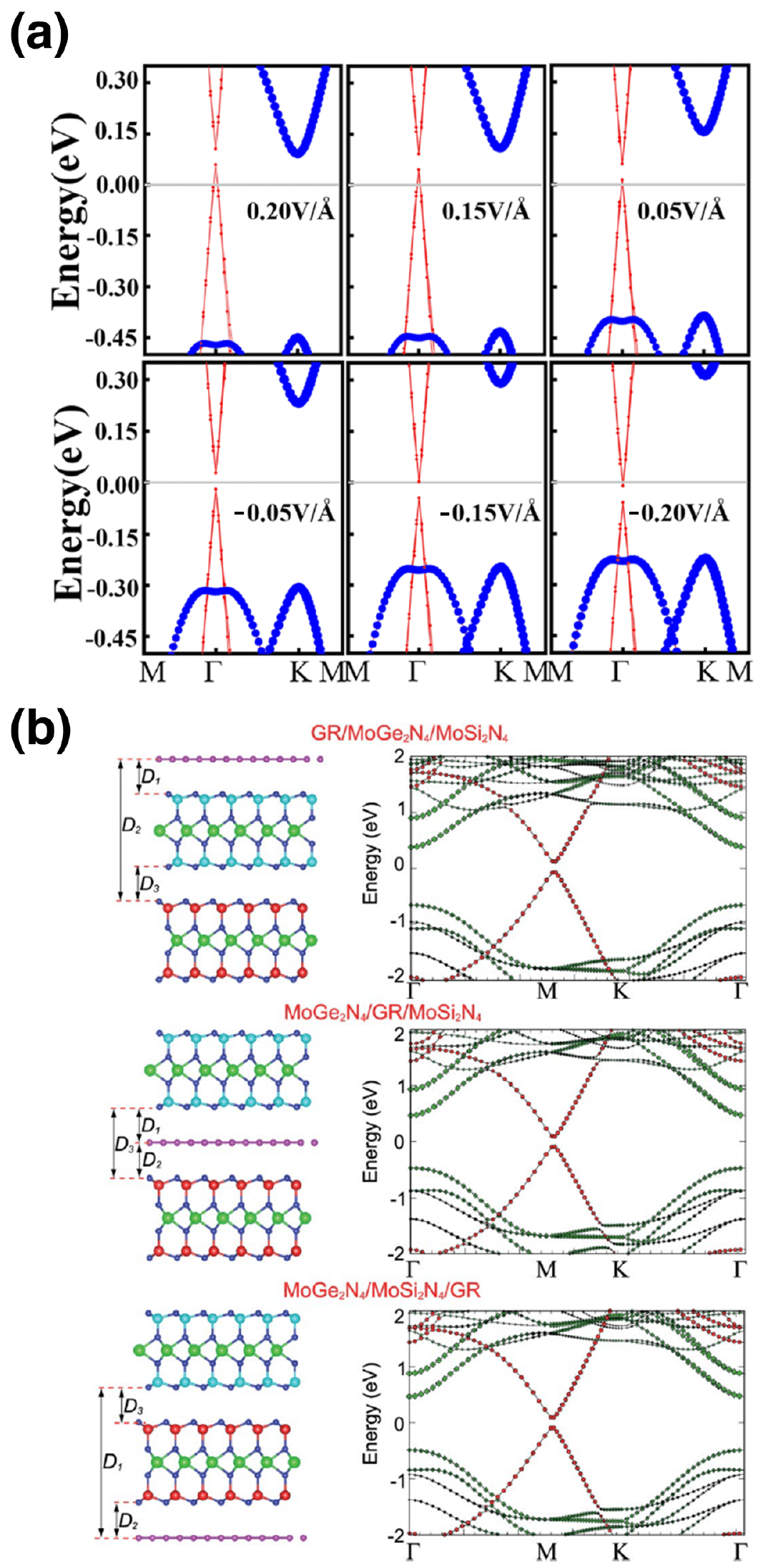}
\caption{\label{Fig11}\textbf{Electronic structures of trilayer MA$_2$Z$_4$ vertical heterostructures.} (a) Electronic band structure of $\mathrm{graphene/BN/MoSi_2As_4}$ under an external electric field in the range of $\pm0.2$ \text{V/\AA}. (b) Electronic band structure of various vertical stacking position of graphene contacting with $\mathrm{MoGe_2N_4/MoSi_2N_4}$. $D_{1}$, $D_{2}$, $D_{3}$ denotes the shortest distance between graphene and $\mathrm{MoGe_2N_4}$, graphene and $\mathrm{MoSi_2N_4}$, $\mathrm{MoGe_2N_4}$ and $\mathrm{MoSi_2N_4}$, respectively. (a) Reproduced with permission from Si $et \ al.$, 2D Mater. \textbf{4}, 015027 (2016), Copyright 2016 IOP Publishing Ltd. (b) Reproduced with permission from Pham, RSC Adv. 11, 28659 (2021), Copyright 2021 Author(s); licensed under a Creative Commons Attribution (CC BY) license.}
\end{figure} 

\begin{figure}[t]
\includegraphics[width=0.4\textwidth]{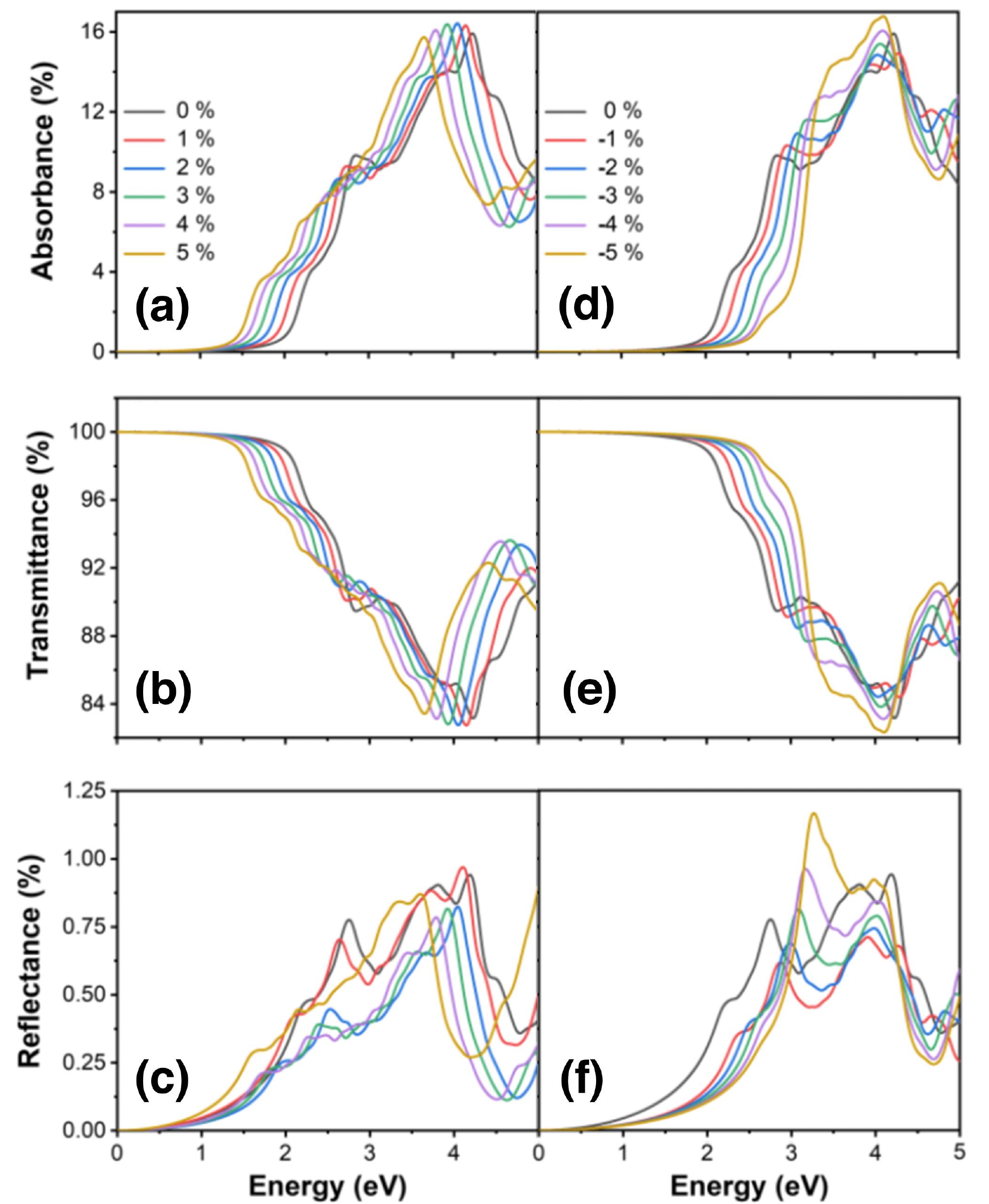}
\caption{\label{Fig12} \textbf{Strain-tunable optical properties of MA$_2$Z$_4$ lateral heterostructures.} Optical spectra of lateral $\mathrm{MoSi_2N_4/WSi_2N_4}$ heterostructure under biaxial strain, showing its (a), (d) Absorbance; (b), (e) Transmittance; (c), (f) Reflectance. Left (Right) panel shows the graphs of tensile (compressive) strain.  (a)-(f) Reproduced with permission from Hussain $et \ al.$, Phys. E: Low-Dimens. \textbf{144}, 115471 (2022), Copyright 2022 Elsevier B.V. All rights reserved.
}
\end{figure}

\begin{figure*}[t]
\includegraphics[width=1.0\textwidth]{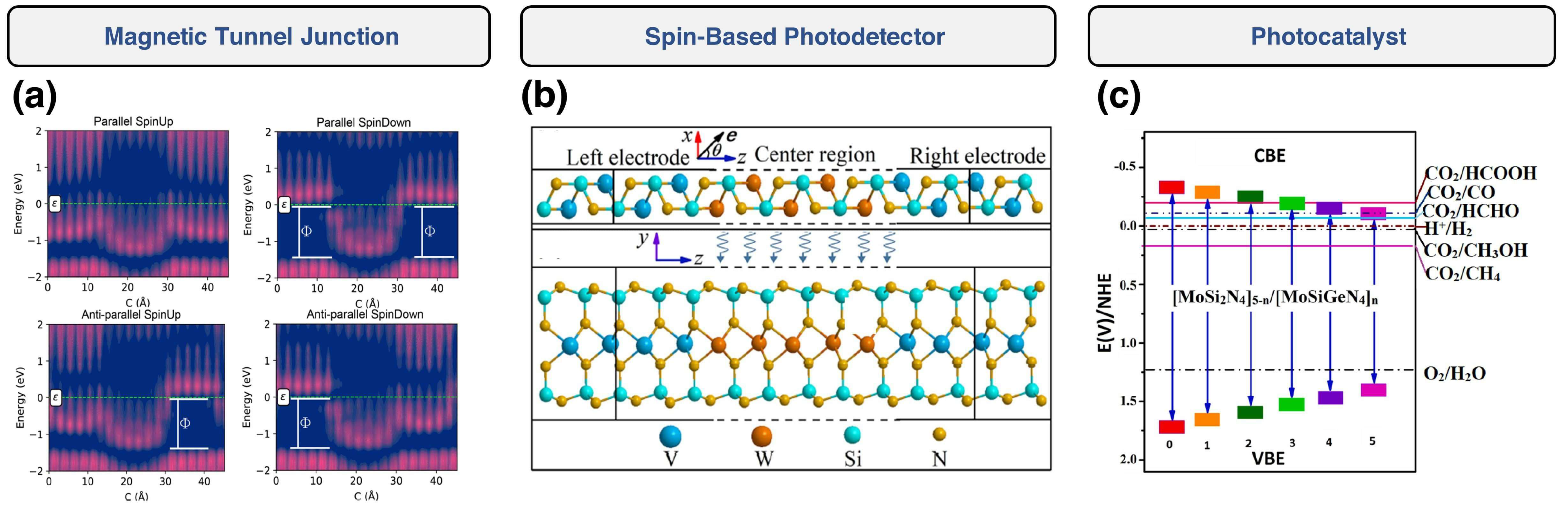}
\caption{\label{Fig13} \textbf{Applications of MA$_2$Z$_4$ lateral heterostructures.} (a) Local spin density of states of $\mathrm{VSi_2N_4/MoSi_2N_4/VSi_2N_4}$ along the device's transport direction. Red(Blue) denotes region of high(low) density. When the metal electrodes are spin-polarized in the same direction (Top Left), there are spin-up states below the Fermi level throughout the whole transport direction, unlike for the cases shown in (Top Right), (Bottom Left), (Bottom Right) where there are spin states missing below the Fermi level. Current is mainly contributed by the spin-up electrons with energies below the Fermi level. (b) Top and side view of the structure of $\mathrm{VSi_2N_4/WSi_2N_4/VSi_2N_4}$ spin-photodetector, where $\mathrm{VSi_2N_4}$ acts as the spin-polarized metal electrode and $\mathrm{WSi_2N_4}$ is used to harvest photons. The structure of $\mathrm{VSi_2N_4/MoSi_2N_4/VSi_2N_4}$ is similar. (c) Band alignment of $\mathrm{(MoSi_2N_4)_{5-n}/(MoSiGeN_4)_n}$ (n = 0,1,2,3,4,5) with respect to the redox potential of water and various reduction potentials of carbon dioxide.
(a) Reproduced from Wu \& Ang, Appl. Phys. Lett. 120, 022401 (2022), with the permission from AIP Publishing. (b) Reproduced with permission from Shu et al., Phys. Rev. Applied 17, 054010 (2022), Copyright 2022 American Physical Society. (c) Reproduced with permission from Mwankemwa et al., Results Phys. 37, 105549 (2022), Copyright 2022 Author(s); licensed under a Creative Commons Attribution (CC BY) license.}
\end{figure*}

\section{\label{Beyond_Heterobilayer} Heterostructures beyond vertical stacking}

\subsection{Trilayered Vertical Heterostructures}

We now review the computational studies of $\mathrm{MA_2Z_4}$-based multilayered vdWHs beyond bilayer morphology. Bandgap engineering of graphene can be achieved in a triple-layered structure of $\mathrm{graphene/BN/MoSi_2As_4}$ in which a small bandgap of either 47 meV or 39 meV in the Dirac point of graphene, depending on the stacking configuration, can be generated \cite{Guo2022_Graphene}. Interestingly, external electric field can induce transitions of the contact types between Schottky and Ohmic contact as shown in Fig. \ref{Fig11}(a). Apart from using different monolayer combinations, the stacking sequence of graphene and MA$_2$Z$_4$ can also be used to engineer the bandgap and overal contact types of the trilayered-heterostructures. When combining graphene with $\mathrm{MoSi_2N_4}$ and $\mathrm{MoGe_2N_4}$ \cite{Pham2021} to form the trilayered-heterostructures of $\mathrm{graphene/MoSi_2N_4/MoGe_2N_4}$, $\mathrm{MoSi_2N_4/graphene/MoGe_2N_4}$ and $\mathrm{MoSi_2N_4/MoGe_2N_4/graphene}$, the bandgap of $\mathrm{MoSi_2N_4/MoGe_2N_4}$ can be reduced and small band gap of (17-46) meV can be opened at the Dirac point of graphene [see Fig.~\ref{Fig11}(b)]. Importantly, the $p$/$n$-type Schottky barrier height between graphene with 
$\mathrm{MoSi_2N_4/MoGe_2N_4}$ bilayer for all three contact positions, are lower than that obtained for graphene contacting with $\mathrm{MoSi_2N_4}$ and $\mathrm{MoGe_2N_4}$ monolayer \cite{Pham2021_TMN}, thus showing that stacking sequence beyond bilayer morphology can be harnessed to effectively tune the electronic and contact properties of vdWHs. 


\subsection{\label{sec:Lateral Heterostructure}Lateral Heterostructure}

Besides $\mathrm{MA_2Z_4}$-based vdWHs, $\mathrm{MA_2Z_4}$-based lateral heterostructures have also been studied recently. To avoid structural instability, covalently bonded lateral heterostructures require small structural mismatch at the contacting interface due to the presence of edge dangling bonds that has to approach each other at appropriate angles. 
An edge contact transistor based on $\mathrm{MoSi_2N_4}$ contacting with 3D Au electrodes reveals that low SBH can be obtained for three different configurations that benefits electrons injection between Au and $\mathrm{MoSi_2N_4}$ \cite{Meng2022}. The metallic members of MA$_2$Z$_4$ family, such as TaSi$_2$N$_4$ and NbSi$_2$N$_4$ can also be used as lateral edge contacts to MoSi$_2$N$_4$ and WSi$_2$N$_4$ semiconducting channels \cite{Qu2023}. Interestingly, the metallic TaSi$_2$N$_4$ and NbSi$_2$N$_4$ are akin to highly $p$-doped MoSi$_2$N$_4$ and WSi$_2$N$_4$ with heavy substitution of Mo and W atoms by Ta and Nb atoms. Because of the narrow-band nature \cite{PhysRevApplied.19.064058} of metallic TaSi$_2$N$_4$ and NbSi$_2$N$_4$ monolayers around the Fermi level, such edge-type electrodes serve as cold source contacts \cite{liu2020switching} that are highly beneficial for achieving steep-slope transistor operation. 
The ultralow subthermionic 10 to 30 mV/decade subthreshold swing achieved in a 12-nm FETs composed of lateral heterostructure $M$Si$_2$N$_4$/$M'$Si$_2$S$_4$ (where $M$ = Mo, W; and $M'$ = Ta, Nb) \cite{Qu2023} suggest the potential usefulness of MA$_2$Z$_4$ MS lateral heterostructures for constructing electronic devices with ultralow power consumption.
Laterally stitched $\mathrm{MoSi_2N_4/WSi_2N_4}$ and $\mathrm{MoSi_2N_4/TiSi_2N_4}$ exhibits strain-tunable optical properties with strong absorbance in the visible light regime [see Fig. \ref{Fig12}(a)-(f)], thus further suggesting their potential usage in solar cell applications \cite{Hussain2022_strain}. 


$\mathrm{VSi_2N_4/MoSi_2N_4/VSi_2N_4}$ forms a magnetic tunnelling junction (MTJ) with a tunnelling magnetoresistance (TMR) mechanism based on the half-metallicity of $\mathrm{VSi_2N_4}$ \cite{Wu2022}. $\mathrm{VSi_2N_4}$ is a ferromagnetic half-metal with  Curie temperature of 500 K which is higher than that of other magnetic 2D monolayers, such as $\mathrm{CrCl_3}$ and $\mathrm{Cr_2Ge_2Te_6}$ (about 50 K). Thus, $\mathrm{VSi_2N_4}$ is potentially useful for spintronic applications at room temperature. Under a biased voltage range of $\pm$100 mV, the TMR which is defined as $TMR=(I_{P}-I_{AP})/I_{AP}\times100\%$ where $I_{P/AP}$ is the charge current across the parallel/anti-parallel polarization between the ferromagnets, is about seven orders of magnitude larger than the previously studied $\mathrm{VS_{2}/MoS_{2}/VS_{2}}$ heterostructure \cite{Zhao2018}. Under the same range of bias voltage, the spin injection efficiency, defined as $\eta_{S}=(I_{\uparrow}-I_{\downarrow})/(I_{\uparrow}+I_{\downarrow} )\times100\%$ where $I_{\uparrow/\downarrow}$ is the up/down polarization of the spin current, is almost 100\% [see Fig. \ref{Fig13}(a) for the local spin density of states along the device's transport direction]. 
Quantum transport simulation shows that the photocurrent of $\mathrm{VSi_2N_4/WSi_2N_4/VSi_2N_4}$ [see Fig. \ref{Fig13}(b) for the device structure] is dependent on the incident and polarization angle \cite{Shu2022}. 
Using appropriate photon energies and polarization angles, the spin-filter and spin-valve effects can be generated, rendering the heterostructure a potential usage as a spin-based photodetector. 

For electrochemical applications, $\mathrm{(MoSi_2N_4)_{5-n}/(MoSiGeN_4)_n}$ \cite{Mwankemwa2022} lateral heterostructure can be used as a photocatalyst for $\mathrm{CO_2}$ reduction [see Fig. \ref{Fig13}(c)]. Changing the chemical composition of the structure via $n$ can effectively tune its optical and photocatalytic properties. In particular, $\mathrm{(MoSi_2N_4)_{0}/(MoSiGeN_4)_{5}}$ exhibits the highest absorption in the visible regime and $\mathrm{(MoSi_2N_4)_{3}/(MoSiGeN_4)_{2}}$ has the lowest activation energy for $\mathrm{CO_2}$ reduction. $\mathrm{(MoSi_2N_4)_{5-n}/(MoSiGeN_4)_n}$ thus offers a promising platform for developing novel sustainable technology. 

\section{\label{outlook} Prospects and Outlooks of MA$_2$Z$_4$ Heterostructures}



Recent advances on the computational design of $\mathrm{MA_2Z_4}$ heterostructures, as reviewed in the previous sections, suggest the potential and strengths of $\mathrm{MA_2Z_4}$ heterostructures in electronics, optoelectronics and renewable energy applications. In this Section, we first provide a concise summary to contrast the strengths of MA$_2$Z$_4$-based heterostructures as compared with other commonly studied 2D materials. This is then followed by a discussion on the prospects and outlooks on the design of MA$_2$Z$_4$ heterostructures moving forward. Recommendations on future research based on emerging computational design techniques, such as data-driven design and computational simulations of the experimental growth of MA$_2$Z$_4$ heterostructures, are presented. 

\subsection{Comparing MA$_2$Z$_4$ heterostructures with other 2D material heterostructures: Strengths and opportunities }

Metal/semiconductor contact interface. Metal-induced gapped states and metallization effect are considerably weaker at the contact interface in MoSi$_2$N$_4$ and WSi$_2$N$_4$-based MS contacts when compared to other 2D semiconductors such as MoS$_2$ in which severe metallization is a commonplace \cite{Gong2014}. Such distinctive behavior arises from the outer Si-N sublayer which acts as a built-in protection layers that protect the band edge electronics states, which predominantly localized in the inner Mo-N core-layer \cite{Wang2021}. 
As the band edges states are well `insulated' from the external contacting metals, the Fermi level pinning effect is substantially weaker, even when contacted by 3D bulk metals that are known to substantially metalize 2D TMDCs, than many other 2D semiconductors. 
The suppression of the Fermi level pinning effect also suggest that the SBH in MoSi$_2$N$_4$ and WSi$_2$N$_4$ MS contacts can be more flexibility engineered via contacting metal electrodes of different Fermi level.

We further note that the unusual built-in sublayer protection -- a distinctive feature not found in TMDC and many other 2D semiconductors such as phosphorene and InSe -- has only been characterized and discussed in MoSi$_2$N$_4$ and WSi$_2$N$_4$ \cite{Wang2021,Liang}. 
Whether such behavior is a general feature \emph{universal} to the MA$_2$Z$_4$ family remains a question that is still open for computational investigation. 
In terms of experimental investigations, transport measurements should be carried out using different work function metals to determine the pinning factors, metal evaporation rates, degree of surface damage and resist residue, and transferred versus deposited contacts. Transfer length method (TLM) and four-point measurements are expected to be useful in determining the contact resistances, which should also be measured at different temperatures to determine the carrier injection mechanisms. We highlight that TLM design should include both short-and long-channel devices, so to separately investigate the contact-limited and the channel-limited transport mechanisms, respectively.

\begin{figure*}
\includegraphics[width=0.85\textwidth]{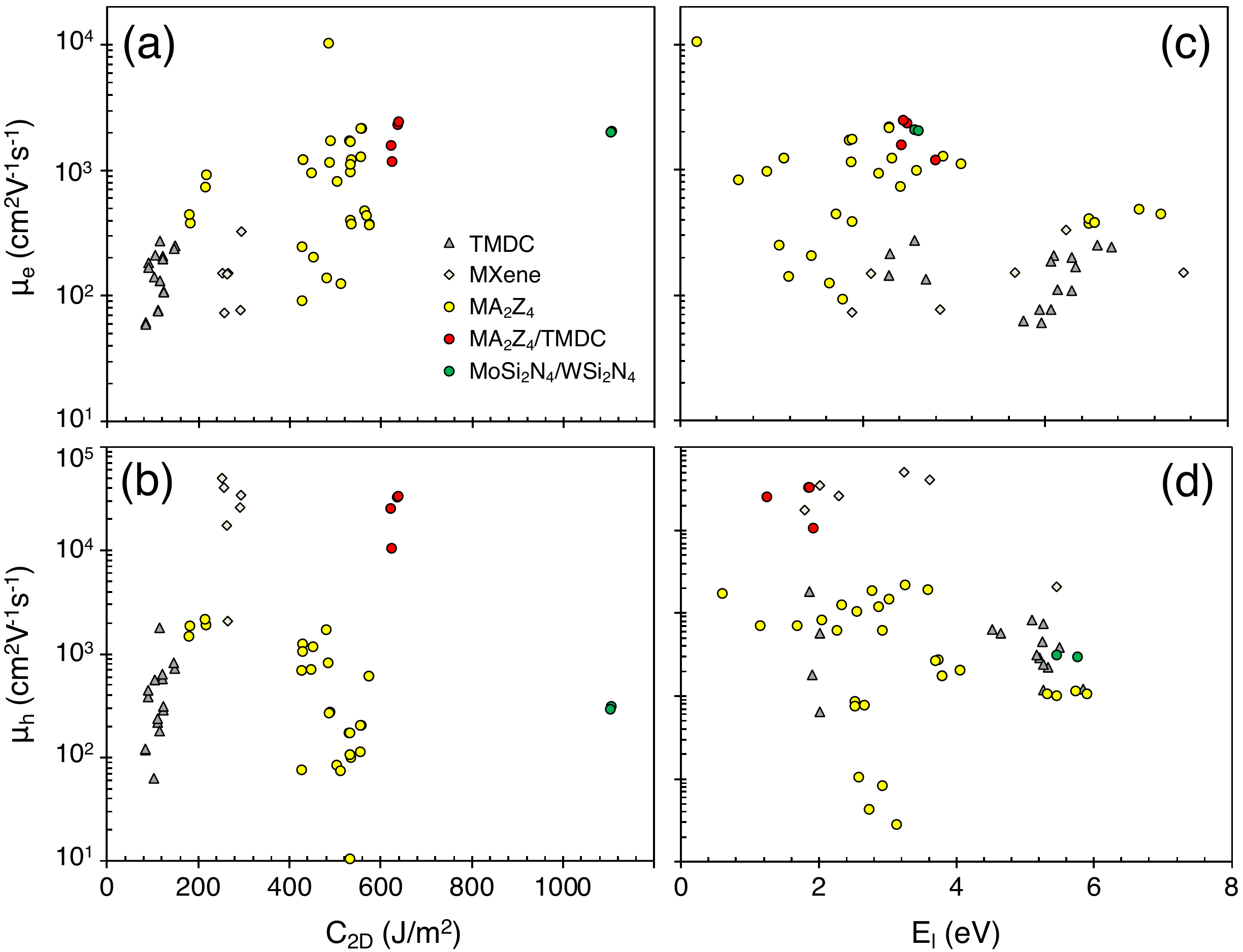}
\caption{\label{Fig14}\textbf{Comparing the carrier mobility of MA$_2$Z$_4$ monolayers, heterostructures with other 2D semiconductors}. Electron mobility as a function of (a) elastic modulus $C_{2D}$ and (b) deformation potential $E_1$. (c) and (d) same as (a) and (b), but for hole carrier mobility. The mobility data are extracted from Refs. \cite{Ren2023,Pei2023,Zhao2022,Yu2021,Rawat2018,Zha2016}. Five families of 2D systems are considered: (i) TMDC monolayers (MX$_2$ where M = Mo, W and X); (ii) MXene (MCO$_2$ where M = Hf, Ti, Zr); (iii) MA$_2$Z$_4$ monolayers (MoSi$_2$N$_4$, WSi$_2$N$_4$, CrSi$_2$N$_4$, HfSi$_2$N$_4$, ZrSi$_2$N$_4$, TiSi$_2$N$_4$, MoGe$_2$N$_4$, MoSi$_2$As$_4$, MoSi$_2$P$_4$, MoSiGeN$_4$, WSiGeN$_4$); (iv) MA$_2$Z$_4$/TMDC heterostructures (WSi$_2$N$_4$/MoSe$_2$, WSi$_2$N$_4$/WSe$_2$); and (v) MoSi$_2$N$_4$/WSi$_2$N$_4$ heterostructures. Both armchair (x) and zigzag (y) directional mobility are included for the monolayers (heterostructures). The full data can be found in the Supplementary Materials. }
\end{figure*}

\textbf{Carrier mobility.} 
Several semiconducting MA$_2$Z$_4$ monolayers, such as the MoSi$_2$N$_4$, WSi$_2$N$_4$, MoGe$_2$N$_4$ and the Janus subfamily of MoSiGeN$_4$, exhibits higher carrier mobility than TMDC monolayers (see Table I).
According to the Bardeen Shockley deformation potential theory (BS-DPT) (see Eq. 4), the carrier mobility is proportional to the elastic modulus $C_\text{2D}$. 
In Fig. \ref{Fig14}, we examine the correlations between the carrier mobility with elastic modulus and deformation potential, for five representative classes of: (i) TMDC monolayers (MX$_2$ where M = Mo, W and X); (ii) MXene (MCO$_2$ where M = Hf, Ti, Zr); (iii) MA$_2$Z$_4$ monolayers (MoSi$_2$N$_4$, WSi$_2$N$_4$, CrSi$_2$N$_4$, HfSi$_2$N$_4$, ZrSi$_2$N$_4$, TiSi$_2$N$_4$, MoGe$_2$N$_4$, MoSi$_2$As$_4$, MoSi$_2$P$_4$, MoSiGeN$_4$, WSiGeN$_4$); (iv) MA$_2$Z$_4$/TMDC heterostructures (WSi$_2$N$_4$/MoSe$_2$, WSi$_2$N$_4$/WSe$_2$); and (v) MoSi$_2$N$_4$/WSi$_2$N$_4$ heterostructure. 
Generally, the electron mobility of MA$_2$Z$_4$ monolayers are higher than that of the other 2D semiconductors [Fig. 14(a)].
Interestingly, the higher electron mobility can be well correlated with their high elastic modulus $C_\text{2D}$ which arises from their thicker lattice where larger force is required to deform the lattice \cite{zhao2022theoretical, Lv2022}.
The $C_\text{2D}$ thus provides a good single descriptor of higher electron carrier mobility in MA$_2$Z$_4$ monolayers and heterostructures.
In contrast, the hole mobility does not strongly correlate with $C_\text{2D}$ [Fig. 14(b)] as the hole mobility is influenced by a more complex interplay between $C_\text{2D}$, $E_1$ and the carrier effective masses. 
For the deformation potential, higher electron and hole mobility can generally be observed in MA$_2$Z$_4$ and their heterostructures with lower $E_1$ [Figs. 14(c) and 14(d)], which arises from the relation of $\mu \propto E_1^{-2}$ according to the BS-DPT in Eq. (4). 

The excellent mechanical properties of MA$_2$Z$_4$ provides another avenue to engineer their carrier mobility. The MA$_2$Z$_4$ family members, such as MoSi$_2$N$_4$ and WSi$_2$N$_4$ , have exceptionally high Young's modulus and mechanical breaking strength \cite{Hong2020, li2021strain} (i.e. 479 and 49 GPa, respectively), which are significantly higher than TMDC monolayers. The exceptional mechanical properties suggest that strain engineering  -- a commonly used method to enhance carrier mobility \cite{mir2020recent,chakraborty2022challenges} -- can potentially be employed to MA$_2$Z$_4$ and their heterostructures over a wider range of mechanical strain to achieve more dramatic mobility tuning.
The potential of achieving high carrier mobility suggests that MA$_2$Z$_4$ monolayer and their heterostructures can potentially be more advantageous than many other 2D material counterparts in applications that requires good transport properties, such as transistors, solar cell and photocatalysts \cite{cheng2019recent}.


We further note that beyond the BS-DPT which includes only the contribution of LA phonon scattering \cite{PhysRev.80.72, PhysRevB.94.235306}, semiclassical transport formalism that rigorously takes into account electron-phonon coupling (EPC) strengths of different phonon modes have been recently applied to calculate the intrinsic phonon-limited mobility of 2D semiconductors \cite{cheng2021intrinsic, PhysRevLett.125.177701, PhysRevB.106.115423}. In particular, BS-DFPT is found to yield inaccurate results for the 2D TMDC family due to the omission of important longitudinal optical (LO) phonon scattering process \cite{cheng2018limits}. We thus propose the EPC-based mobility formalism to be carried out to characterize the intrinsic carrier mobility of MA$_2$Z$_4$ monolayers and heterostructures. 
Such investigations shall yield insights on the phonon modes that limit the carrier conduction of MA$_2$Z$_4$ family. 
Furthermore, recent high-throughput screening of 2D semiconductor with high room-temperature mobility have demonstrated a simpler carrier mobility calculation that includes the effect of both LA and LO phonons, which can be expressed via a Matthiessen’s rule, $\mu^{-1}_\text{intrinsic} = \mu^{-1}_\text{BS-DPT} + \mu^{-1}_\text{F}$, where $\mu_\text{BS-DPT}$ and $\mu_\text{F}$ are the LA phonon-limited carrier mobility calculated via BS-DPT and the LO phonon-limited carrier mobility due to Frohlich scattering \cite{zhang2023two}. 
The simplified nature of $\mu^{-1}_\text{intrinsic}$ may provide an accelerated method to characterize the intrinsic carrier mobility for the expansive family of MA$_2$Z$_4$ monolayers and their heterostructures beyond the potentially oversimplified BS-DPT description.

\textbf{Optical absorption and solar energy conversion performance.} Many MA$_2$Z$_4$-based vdWHs are predicted to have high optical absorption ability in the visible and UV regime, which are aspects also found in TMDC-based vdWHs \cite{Chang2014, Cui2019, Peng2016, Wang2014_gC3N4/MoS2, Song2021, Lan2016, Qiao2022, Cen2023, Pandey2020} though more notably in the visible regime for the latter. The predicted PCE values of a few MA$_2$Z$_4$-based vdWHs have been simulated \cite{He2022,Tho2022,Zhang2023}, showing high PCE reaching $>$ 20\%  \cite{Linghu2018}. The competitive PCE values, coupled with the strong optical absorption properties and high carrier mobility of MA$_2$Z$_4$-based vdWHs, suggest that these materials are promising for photovoltaic applications. 
Furthermore, the generally larger thickness of Janus-MA$_2$Z$_4$ monolayers, as compared to Janus-TMDC, is expected to produce a larger spatial separation of the photoexcited electron-hole pairs, which is beneficial for preventing the recombination loss in solar cell application.
We further note that the studies of MA$_2$Z$_4$ vdWH excitonic solar cell mostly estimate the \emph{maximal} PCE assuming the incident solar spectrum is 100\% absorbed -- a feature not straightforwardly achievable in practical implementations.
We note that the simulation of excitonic solar cell PCE that takes into account the actual optical absorption properties of the MA$_2$Z$_4$ vdWHs, instead of assuming 100\% absorption, have yet to be conducted. 
We propose such calculations to be carried out, so as to facilitate a more realistic characterizations of the solar energy conversion performance of MA$_2$Z$_4$ vdWHs.
In particular, the G$_0$W$_0$-BSE optical calculations should be used to determine more accurate band gap values and the excitonic effects are better captured in the modelling \cite{Leng2016}.


\subsection{Prospects and outlook of $\mathrm{MA_2Z_4}$-based heterostructures}

As the initial `gold rush' of MA$_2$Z$_4$ is settling down, it is now important to list the bottlenecks and challenges as well as to re-strategize towards potential new research frontiers of $\mathrm{MA_2Z_4}$ monolayers and their heterostructures. Below we summarize several key prospects of $\mathrm{MA_2Z_4}$ heterostructures that are potentially useful in driving the next-phase of $\mathrm{MA_2Z_4}$ heterostructure research.

\textbf{Prospect 1. Experimental realization of MA$_2$Z$_4$ and heterostructures.} Although a large variety of $\mathrm{MA_2Z_4}$ monolayers have been predicted and studied computationally \cite{Yin2022}, only $\mathrm{MoSi_2N_4}$ and $\mathrm{WSi_2N_4}$ have been experimentally synthesized thus far. The progress of synthesizing the broader MA$_2$Z$_4$ monolayer family is further impeded by the lack of bulk parent, thus making mechanical exfoliation -- one of the primary and relatively straightforward tools in isolating 2D monolayers \cite{yi2015review} -- impossible. 
The lack of experimental realization of the monolayers represent a major roadblock in the quests of transforming MA$_2$Z$_4$ heterostructure from theoretical research into practical device implementations \cite{zhang20222d}. 
Experimental efforts focusing on synthesizing MA$_2$Z$_4$ monolayers beyond $\mathrm{MoSi_2N_4}$ and $\mathrm{WSi_2N_4}$, so to confirm previous theoretical predictions and to narrow down the \emph{experimentally feasible} candidate pool for subsequent theoretical investigations.
Recently, high-throughput experimental fabrication and characterization methods have been employed for nanomaterials discovery and design \cite{10.1063/1.4977487, hattrick2015materials}, including 2D material heterostructures \cite{maggini20212d}. 
Leveraging on highly-automated and highly-parallel fabrication as well as measurement workflows \cite{liu2019high}, we anticipate high-throughput experiments to provide a viable route towards the synthesis of MA$_2$Z$_4$ monolayers and heterostructures.
In particular, recent demonstration of rapid CVD growth of MoS$_2$ wafer suggests \cite{seol2020high} that similar method could potentially be developed to navigate the huge combinatorial space that covers various growth parameters, such as the substrate species, precursors species and concentrations, temperature and precursor gas flow rate that sensitively influence the growth outcomes, thus enabling the rapid identification the growth recipe for a specific MA$_2$Z$_4$. 
This aspect also highlights the importance of computational simulations in providing guidance on the material candidates and growth mechanisms (see also the \textbf{Prospects 2 and 3} below), so that the potentially tremendous experimental efforts needed to realize a specific functional MA$_2$Z$_4$ can be narrowed down towards a smaller pool of more promising candidates.  

We further note that a peculiar `super-thick' MoSi$_2$N$_4$(MoN)$_n$ monolayer, formed by the intercalation of $n$ submonolayers of MoN with Si-N outer submonolayers, has been very recently demonstrated experimentally \cite{NSR}. The super-thick synthetic MoSi$_2$N$_4$(MoN)$_n$ monolayers hinted that such intercalation architecture may be generalized to other members MA$_2$Z$_4$ family to yield a new family of MA$_2$Z$_4$(MZ)$_n$ intercalated monolayers or even to render a Janus morphology of MA$_2$Z$_4$(M'Z')$_n$ (where M $\neq$ M' and Z $\neq$ Z'). 
We thus anticipate more surprises to emerge from the `super-thick' MA$_2$Z$_4$(MZ)$_n$ and their Janus counterpart MA$_2$Z$_4$(M'Z')$_n$, which shall further expand the potential usefulness of MA$_2$Z$_4$ family. 

\textbf{Prospect 2. Computational simulation of MA$_2$Z$_4$ and heterostructure growth mechanisms.} 
In relevance to the experimental challenges mentioned above, first-principles-based computational simulations provide powerful predictive tools to understand the growth mechanism \cite{https://doi.org/10.1002/adts.201800085, gao2018computational, dong2021theoretical} of MA$_2$Z$_4$ monolayers and heterostructures. As DFT generates only the ground state properties, DFT alone is insufficient to model the \emph{dynamical} material growth process. Dynamical methods, such as kinetic Monte Carlo (KMC), classical molecular dynamics (MD) or the computationally more expensive AIMD are needed combined with DFT to model the growth mechanisms of 2D materials and heterostructures \cite{bhowmik2022chemical}. Various DFT-based MD and KMC simulations have been employed to understand the CVD growth and nucleation processes of 2D materials, such as graphene \cite{zhang2011first}, MoS$_2$ \cite{lei2022salt}, WSe$_2$ \cite{nie2016first}.
Such DFT-based simulations shed lights on the interactions between precursors and substrates \cite{gao2011graphene}, edge termination \cite{zhang2014role} and vertical heterostructure growth mechanisms \cite{zhang2016twinned}, which are useful for the experimental CVD synthesis of 2D materials and heterostructures.
Critically, as MA$_2$Z$_4$ is \emph{synthetic} without bulk parents, exfoliation synthesis methods \cite{islam2022exfoliation} which are widely used in producing high-quality 2D materials are not compatible for the MA$_2$Z$_4$ family. This aspect emphasizes the importance conducting DFT-based simulations to extract the underlying physics and chemistry of the CVD growth mechanisms, so to yield critical insights on the optimal growth conditions, such as substrate, precursors and temperature, that are crucial for achieving high-quality and large-area samples critically needed for constructing heterostructures. 

\begin{figure*}
\includegraphics[width=1\textwidth]{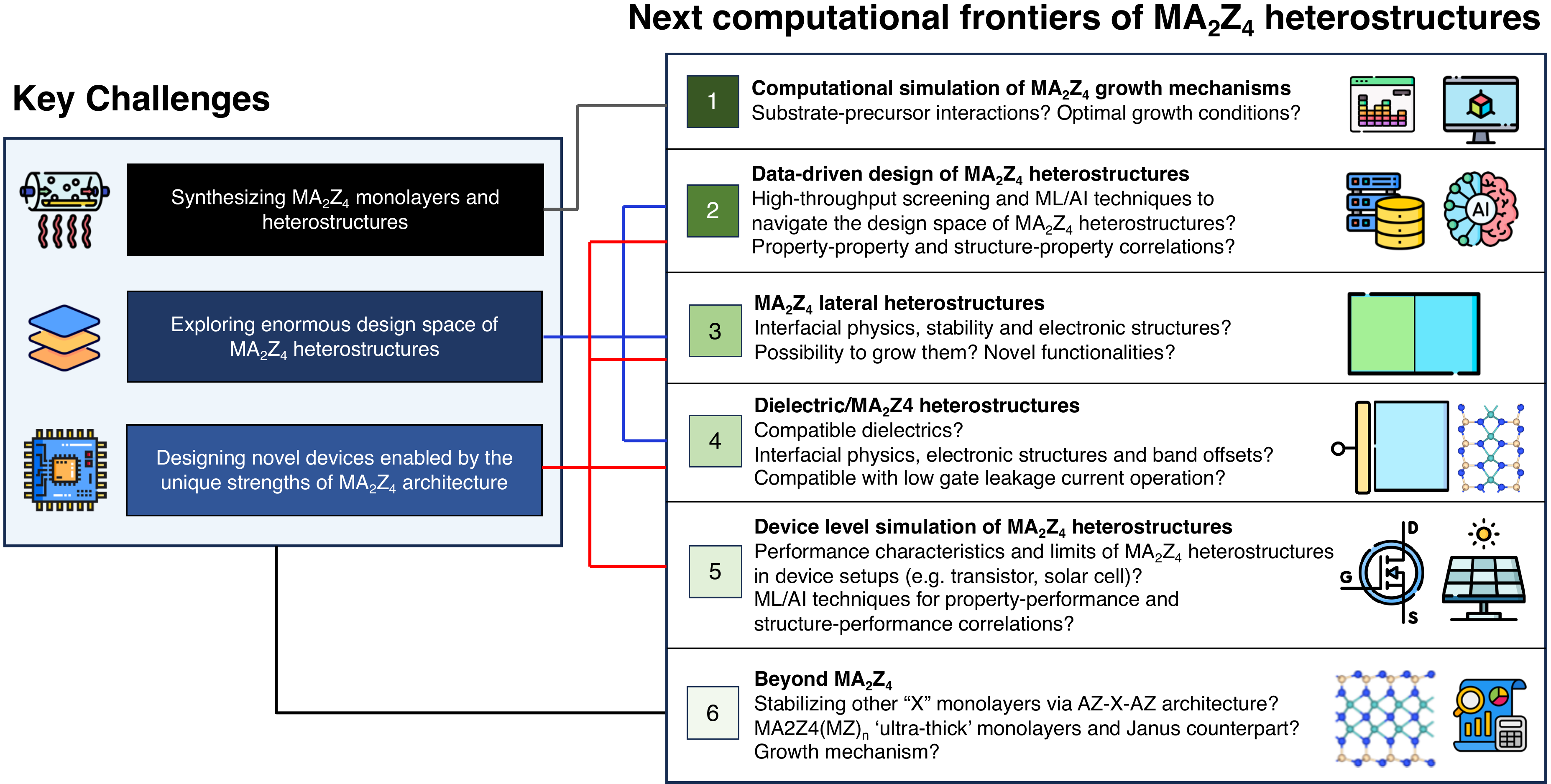}
\caption{\label{Fig15}\textbf{Summary on the challenges and proposed new computational frontiers in MA$_2$Z$_4$ heterostructure research}. The three key challenges of MA$_2$Z$_4$ heterostructure research, namely (i) synthesis of MA$_2$Z$_4$ monolayers and heterostructures; (ii) Navigating through the enormous design space of MA$_2$Z$_4$ heterostructures; and (iii) Designing devices with high performance and novel device architectures that leverage on the unique material strengths of MA$_2$Z$_4$ and their heterostructures, are proposed to be addressed via six computational Frontiers. Frontier 1 provides insights on the growth mechanisms useful for addressing Challenge (i). The Challenge (ii) can be addressed by the data-driven approach and various heterostructure architectures targeted in Frontiers 2 to 4. The Challenge (iii) can be similarly addressed by Frontiers 2 to 4, in combination with the device-level studies outlined in Frontier 5. Finally, Frontier 6 is expected to enrich and expand the family of MA$_2$Z$_4$ with similar intercalation architecture. }
\end{figure*}

\textbf{Prospect 3. Data-driven approach of heterostructures design and discovery: High-throughput screening and machine learning models.} High-throughput material screening offers an emerging route for the discovery and design of functional 2D materials \cite{mounet2018two}, thanks to the recent advancements of supercomputing facilities and large-scale material databases, such as 2DMatPedia \cite{zhou20192dmatpedia} and C2DB \cite{haastrup2018computational, gjerding2021recent}. 
High-throughput material screening typically starts with a large material database containing hundreds or thousands of initial candidates. The candidate pool is then subsequently reduced based on the sequential implementation of \emph{filters} with increasing computational cost, such as composing elements, band gap, lattice symmetry, mobility, mechanical/thermodynamical/dynamical stability and so on \cite{zhang2019high, shen2022high}. 
The screening process eventually produces a smaller pool of material candidates, in which computationally costly can then be performed to understand the performance characteristics and limits for specific target function(s). 
Based on high-throughput screening methods, a large variety of 2D materials with specific functionalities, such as ferroelectricity \cite{kruse2023two}, ferroelasticity \cite{PhysRevLett.129.047602}, excitonic solar cell \cite{linghu2018high}, topological insulator \cite{olsen2019discovering}, high-photocatalysts \cite{singh2015computational} and superconductivity \cite{wines2023high}, have been catalogued computationally.
High-throughput screening has also been employed to the discovery 2D/2D heterostructures, for examples, for solar cell and electrodes applications. 
However, it should be noted that due to the enormous design space of 2D material heterostructures, the procedures is often limited by: (i) the use of oversimplified methods, such as the Anderson rule which omits the important interfacial interactions between the two contacting surfaces; (ii) a smaller initial candidate pool for screening to reduce the computational cost, which may not fully capture high-performing candidates that lie beyond the smaller initial candidate pool.

High-throughput screening combined with machine learning (ML) methods provide an alternatively route towards data-driven material design and discovery. 
For example, the band alignment types of 2D/2D heterostructures have been investigated using ML models in recent works \cite{Choudhary_2023,Zheng2021,Willhelm2022}. Explicit DFT calculations is first conducted to form the training and testing dataset. An appropriate ML model is first trained using monolayer properties, such as the band edge positions, which are computationally less costly or are readily available in existing material databases. The validity ML model prediction is then verified using the test dataset. Such ML-based method generates useful property-property correlation that streamlines the discovery of new heterostructures and accelerates the prediction of their performance characteristics wihtout the need of costly explicit DFT calculations. 
Furthermore, ML models can also be used to extract structure-property correlations, which are particularly useful in understanding how structural defects are correlated with the material properties \cite{huang2023unveiling}.

The emergence of data-driven material discovery and design approaches, either based solely on high-throughput `brute force' method or a combination of high-throughput and ML methods for property-property and structure-property correlations extractions, shall provide a feasible avenue to tame the enormous design space of MA$_2$Z$_4$ heterostructures. 
We anticipate high-throughput and ML methods to shed light on the following critical unresolved issues that may impact the performance of electronics and optoelectronics devices based on MA$_2$Z$_4$ and their heterostructures: 

(1) \emph{Interfacial band alignment} - The vdW interactions in 2D/2D heterostructures are beneficial in preserving the brand structures of the constituent monolayers, but interfacial charge transfer is still a commonplace in most contact heterostructures. Such transfer gives rise to a built-in electric dipole potential difference $\Delta V$, which shifts the relative band alignments between the two constituent monolayers and causes them to deviate from the Anderson rule \cite{besse2021beyond}. Currently, the determination of $\Delta V$ requires explicit DFT calculations that are computationally costly. Simple analytical model of $\Delta V$ that may circumvent DFT calculations are challenging to construct due to the intricate and complex interactions at the contact interface. Here data-driven approach may provide an alternative route to predict $\Delta V$ without explicit DFT calculations. In particular, ML models can be used to extract generalizable property-property correlations of $\Delta V$, i.e. $\Delta V_\text{ML}$, based on dataset generated via high-throughput heterostructure DFT calculations, thus resurrecting the Anderson rule using $\Delta V_\text{ML}$ as a correction term. Similar data-driven approach can also be implemented in MS contact heterostructures for the rapid identification of SBH values.

(2) \emph{Interfacial defects} - Defects are known to substantially modulate the interfacial properties of 2D materials \cite{bussolotti2021impact}. The diverse choices of atomic site substitution in MA$_2$Z$_4$, together with the large combinatorial space of contact heterostructures (i.e. stacking orientations and material combinations \cite{boland2022computational}) shall be benefited from the high-throughput and ML approach.

(3) \emph{Heterostructure device performance} - A more ambitious application of data-driven method is the extraction of direct correlations between structure, property and \emph{device-level performance} (such as carrier mobility and solar cell power conversion efficiency), namely the \emph{structure-performance} and \emph{property-performance} correlations. ML techniques have been previously demonstrated in the modelling of transistor, such as compact modelling \cite{carrillo2019machine, woo2022machine} and failure analysis \cite{chatterjee2021machine, genssler2023modeling}, chiefly at the device level. How the data-driven approach can be utilized to link the material-level properties and device-level performance characteristics, which are otherwise challenging and overly complex using physics-based models, remain an open question thus far. Such a data-driven correlations extraction workflow is expected to involve the generation of large dataset using high-throughput DFT simulations and additional device-level simulations (also see \textbf{Prospect 4} below), before employing ML model to extract practically useful structure-performance and property-performance correlations. 
We anticipate such data-driven extractions of structure-performance and property-performance correlations to provide a powerful platform for accelerating the design of MA$_2$Z$_4$ heterostructure devices, which can potentially be further generalized to other classes of 2D materials and devices. 

\textcolor{black}{Finally, we note that the constructions of ML models require sufficiently large data sets. Due to the limited availability of experimental data, computationally generated data shall play an important role for ML model constructions for MA$_2$Z$_4$ heterostructure design. Such computational data sets can be generated via automated high-throughput calculation workflow that scan through existing large material databases \cite{zhou20192dmatpedia, gjerding2021recent}. Apart from performing explicit high-throughput DFT calculations for data generation, specialized \emph{small data ML} methods, such as imbalance data learning, transfer learning, and active learning, can also be used to tackle the challenges of limited data sets, and such methods have been widely employed in material science research \cite{xu2023small}.}

\textbf{Prospect 4. Device-level simulation and novel device architectures.} Although $\mathrm{MA_2Z_4}$-based vdWHs have been predicted to exhibit myriads of promising capabilities for device applications, most of the theoretical predictions report only the basic physical quantities (e.g. band alignment, absorption coefficients and Schottky barrier heights), but the performance at the \emph{device level} (such as transistor, photodetector and solar cell) remains largely unexplored thus far. Such device-level simulations are urgently needed to convincingly establish the technological and industrial relevance of $\mathrm{MA_2Z_4}$-based heterostructures. Furthermore, \emph{novel functional devices}, beyond current or conventional device architecture, that are uniquely enabled by the material properties of MA$_2$Z$_4$-based vdWHs has rarely been explored, such as investigation into the nanodiode architecture based on graphene/MoSi$_2$N$_4$/NbS$_2$. We thus expect the computational device simulations and novel device designs to be the next key areas to explore. The simulation results can also be incorporated with high-throughput and ML methods as discussed in \textbf{Prospect 4} above to yield useful device design rules. 

\textbf{Prospect 5. Dielectric/MA$_2$Z$_4$ heterostructures.}
Recent investigations of MA$_2$Z$_4$ heterostructures focus mainly on MS and SS contact types. Interfacing MA$_2$Z$_4$ with \emph{dielectric} materials, such as 3D bulk dielectrics of SiO$_2$ and HfO$_2$ \cite{yang2023gate} as well as the emerging 2D counterparts \cite{osanloo2021identification, vandenberghe2023two}, such as hBN \cite{laturia2018dielectric} and Bi$_2$SeO$_5$ \cite{zhang2022single}, has yet to be systematically explored thus far. 
As dielectric/semiconductor interface is an indispensable component of FET \cite{yang2023gate}, the important contact class of \emph{dielectric/MA$_2$Z$_4$ heterostructures} is expected to yield important knowledge on the feasibility of MA$_2$Z$_4$ in electronics applications. 
Importantly, an appropriately engineered dielectric interface, such as the use of high-k dielectrics \cite{jena2007enhancement} and dielectric substrate roughness \cite{ng2022improving, liu2019crested}, can lead to mobility enhancement in 2D semiconductor.
Scrutinizing the interfacial contact physics of dielectric/MA$_2$Z$_4$ heterostructures is thus expected to yield insights on the following open research questions:
(i) Does dielectric/MA$_2$Z$_4$ interface exhibit ideal band alignment with sufficiently large conduction and/or valance band offsets that are desirable in suppressing the gate leakage current?
(ii) In 2D semiconductors like TMDC, the pristine surfaces have proved to be a double edge sword, useful in preserving carrier mobility down to the monolayer limit, but also severely limiting the dielectric integration via industrial techniques such as atomic layer deposition as precursor nucleation is inhibited. Will MA$_2$Z$_4$ face similar challenges in dielectric integration?
(iii) Can high-k dielectric be used to enhance the carrier mobility in an appropriately designed dielectric/MA$_2$Z$_4$ interface?
We propose dielectric/MA$_2$Z$_4$ contact heterostructures to be one of the next frontiers of MA$_2$Z$_4$ heterostructure research.

\textbf{Prospect 6. MA$_2$Z$_4$ lateral heterostructures.} Laterally stitched $\mathrm{MA_2Z_4}$-based heterostructures remains a largely unexplored area. Such heterostructure architecture can be beneficial for lowering contact resistance when compared to vertical heterostructures which are often plagued by the presence of a vdW tunneling barrier \cite{Chakraborty2022}. Furthermore, as current state-of-the-art methods of degenerately doping the source and drain contacts of the silicon FET to lower contact resistance poses a huge challenge to map onto atomically thin materials, the growth of lateral MS heterostructures can provide a route to seamlessly attach metallic and semiconducting $\mathrm{MA_2Z_4}$ monolayers together, thus circumventing the need of degenerately doping the source and drain contact. 

In fact, the design space of lateral heterostructures is comparable, if not larger, than that of the vertical heterostructures. The construction of lateral heterostructure involves: (1) materials combination; (2) stitching configuration; and (3) defects. Furthermore, because the monolayers are covalently bonded with potentially larger lattice distortion \cite{zhao2018growth}, in contrast to the weak vdW coupling of vertical heterostructures, the structural stability can be a more severe issue for lateral heterostructures. 
This aspect necessitates more a stringent stability verification via phonon spectrum, AIMD and elastic calculations, thus suggesting a higher computational cost. 
Importantly, unlike vertical heterostructure stacking, lateral stitching of two 2D materials cannot be achieved via mechanical transfer \cite{wang2019recent, ling2016parallel}. The growth mechanism of lateral MA$_2$Z$_4$ heterostructures (related to \textbf{Prospect 2} above) is thus expected to serve as a key practicality indicator that is useful to avoid the `pseudo-discovery' of spurious lateral heterostructure that cannot be realized experimentally. 
We thus expect computational studies of lateral heterostructures to be another vibrant area of MA$_2$Z$_4$ heterostructures research.

\textbf{Prospect 7. Generalizing the intercalation architecture for designing novel AZ-$X$-AZ monolayer.} 
One of the original motivations of synthesizing MoSi$_2$N$_4$ and WSi$_2$N$_4$ monolayers is to demonstrate that the unstable transition metal nitride monolayers (i.e. MoN$_2$ and WN$_2$) can be stabilized using Si-N outer layers based on an intercalation architecture with SiN-MoN$_2$-SiN morphology \cite{Hong2020}. 
Such approach motivates the discovery of the broader MA$_2$Z$_4$ family \cite{Wang2021_MA2Z4} with the morphology of AZ-MZ$_2$-AZ.
Very recently, such intercalation architecture has been applied to graphene, yielding CSiN monolayer with the same morphology, i.e. SiN-graphene-SiN \cite{yan2023triggering}.
This exploration opens up the question of whether such intercalation approach can be generalized to stabilize other monolayers that are relatively unstable or prone to rapid degradation under ambient conditions, such as the  \emph{sister structures} of graphene \cite{balendhran2015elemental} including silicene \cite{zhao2016rise}, germanene \cite{acun2015germanene}, stanene \cite{zhu2015epitaxial} and phosphorene \cite{liu2015semiconducting}. 
We expect the computational exploration of AZ-$X$-AZ monolayers, where $X$ is an inherently unstable monolayer or monolayer prone to rapid degradation, and how such synthetic monolayers can be grown via CVD method (related to \textbf{Prospect 2} above) to represent another exciting frontiers to explore beyond the `conventional' MA$_2$Z$_4$ family.

\subsection{Summary of challenges and new computational frontiers of MA$_2$Z$_4$ heterostructure.}
We summarize the challenges and potential new computational/theoretical research frontiers for the MA$_2$Z$_4$ heterostructure outlined in Sections VI.A and VI.B above in Fig. \ref{Fig15}. Computational tools, such as DFT, MD and quantum transport device simulations, can potentially address the three key challenges of MA$_2$Z$_4$ heterostructures, namely (i) synthesizing MA$_2$Z$_4$ monolayers beyond MoSi$_2$N$_4$ and WSi$_2$N$_4$ monolayers \cite{Hong2020} as well as their heterostructures; (ii) how to efficiently navigate through the enormously large design space of MA$_2$Z$_4$ heterostructures; and (iii) constructing a comprehensive `catalogue' of performance characterization and limits for the expansive family of MA$_2$Z$_4$ heterosturctures, via six research directions which remains largely incomplete thus far (see Fig. 14). For instance, the growth mechanisms (Frontier 1) shall yield insights that can accelerate the experimental realization of MA$_2$Z$_4$ monolayers and heterostructures [i.e. Challenge (i)]. Data-driven approach and the investigations of lateral heterostructure as well as dielectric/MA$_2$Z$_4$ contacts (Frontiers 2 to 4) shall enable the interfacial physics, properties and the design guidelines of MA$_2$Z$_4$ heterostructures to be better understood, thus closing some of the knowledge gap related to Challenge (ii). Frontiers 2 to 4, combined with device-level simulations, shall enable the performance characterizations and limits to be catalogued for conventional devices (such as sub-10-nm transistor and solar cell) as well as novel functional devices (such as spintronic and valleytronic devices), thus addressing Challenge (iii). Finally, the computational material discovery and design outlined in Frontier 6 may yield (1) new subfamily that exhibits a similar intercalation morphology of AZ-$X$-AZ; (ii) ultra-thick monolayers of MA$_2$Z$_4$(MZ)$_n$; and (iii) various Janus variations of (i) and (ii), which shall further enrich the field of MA$_2$Z$_4$ heterostructure.

\section{\label{conclusion}Conclusion}

In summary, the interfacial properties and contact types of MA$_2$Z$_4$-based heterostructures have been reviewed. 
MA$_2$Z$_4$-based heterostructures provide a versatile platform to explore a myriad of next-generation devices that are critically needed to resolve some of the ongoing challenges, such as beyond silicon nanoelectronics and clean energy technologies. 
For $\mathrm{MA_2Z_4}$-based MS contacts, the tunnelling probability and specific resistivity across the MS vdWHs are more favourable when the $\mathrm{MA_2Z_4}$ monolayer is in contact with 3D metals. On the other hand, $\mathrm{MA_2Z_4}$-based SS vdWHs can be classified into either Type-I, Type-II, Type-III or Type-H, whereby Type-II category has been extensively studied due to the solar cell and photocatalytic ability of these materials. 
Type-II $\mathrm{MA_2Z_4}$-based SS vdWHs with a strong intrinsic electric field not only promote the efficient spatial separation of electrons and holes, but also surmounts the minimum 1.23 eV bandgap requirement for water splitting through either partaking in the Z-scheme photocatalytic process or aided by the additional potential provided by the vacuum level offset at the different monolayers. 
The photocatalytic abilities of $\mathrm{MA_2Z_4}$-based heterostructures further suggest their capabilities in high-performance renewable energy applications which are critically needed to achieve sustainable future development. Finally, we discussed the challenges and potential new frontiers of $\mathrm{MA_2Z_4}$-based heterostructures. We hope that the prospects presented in this work can provide potential new directions on the computational and theoretical simulations of $\mathrm{MA_2Z_4}$ heterostructures that may potentially lead to fruitful results after the initial `gold rush' of $\mathrm{MA_2Z_4}$ research.

\section*{Supplementary Material}
See the supplementary material for data plotted in Fig. \ref{Fig14}.

\section*{\label{sec:Acknowledgement}Acknowledgement}
This work is funded by the Singapore Ministry of Education (MOE) Academic Research Fund (AcRF) Tier 2 Grant (MOE-T2EP50221-0019) and SUTD-ZJU IDEA Visiting Professor Grant (SUTD-ZJU (VP) 202001). Figure 15 has been designed using images from Flaticon.com.

\section*{\label{sec:Declaration}Author Declarations}
\subsection*{Conflicts of Interest}
The authors have no conflicts to disclose. 

\subsection*{Author Contributions}
\textbf{Che Chen Tho}: Data curation (lead); Formal analysis (lead);
Investigation (equal); Writing – original draft (equal); Visualization (lead); Writing – review \& editing (supporting);
\textbf{San-Dong Guo}: Writing – review \& editing (supporting);
\textbf{Shi-Jun Liang}: Writing – review \& editing (supporting);
\textbf{Wee Liat Ong}: Writing – review \& editing (supporting);
\textbf{Chit Siong Lau}: Writing – review \& editing (supporting);
\textbf{Liemao Cao}: Writing – review \& editing (supporting);
\textbf{Guangzhao Wang}: Conceptualization (supporting); Data curation (supporting); Writing – original draft (supporting); Writing – review \& editing (supporting);
\textbf{Yee Sin Ang}: Conceptualization (lead); Investigation (equal); Formal analysis (supporting); Supervision (lead); Funding acquisition (lead); Visualization (supporting); Writing – original draft (equal); Writing – review \& editing (lead).


\section*{Data Availability Statement}

Data sharing is not applicable to this article as no new data were
created in this study.

\providecommand{\noopsort}[1]{}\providecommand{\singleletter}[1]{#1}%

\end{document}